\documentclass[11pt]{article}		%

\usepackage[sort&compress,numbers]{natbib}
\usepackage{amssymb,graphicx,lineno}
\usepackage{rotating}
\usepackage{enumerate}
\usepackage{setspace}
\usepackage{multirow, color}
\usepackage[bookmarks=false, pagebackref, colorlinks, allcolors=blue]{hyperref}
\usepackage{siunitx}
\usepackage{listings}
\usepackage{appendix}
\usepackage{setspace}
\usepackage{booktabs,subcaption,longtable,threeparttable}
\usepackage{amsmath}
\usepackage{hyperref}
\usepackage{authblk}
\usepackage{mhchem}

\begin{document}



\title{Influence of ensemble boundary conditions  (thermostat and barostat) on the deformation of amorphous polyethylene by molecular dynamics}

\author[1,2]{M.A. Tschopp}
\author[2]{J.L. Bouvard}
\author[3]{D.K. Ward}
\author[2,4]{D.J. Bammann}
\author[2,4]{M.F. Horstemeyer}

\affil[1]{Dynamic Research Corporation, (DRC), High Performance Technologies Group at the U.S.~Army Research Laboratory, Aberdeen Proving Ground, MD 21005}
\affil[2]{Center for Advanced Vehicular Systems, Mississippi State University, Starkville, MS 39759}
\affil[3]{Sandia National Laboratory, Livermore, CA 94551}
\affil[4]{Department of Mechanical Engineering, Mississippi State University, Starkville, MS 39759}

\maketitle


\begin{abstract}

Molecular dynamics simulations are increasingly being used to investigate the structural evolution of polymers during mechanical deformation, but relatively few studies focus on the influence of boundary conditions on this evolution, in particular the dissipation of both heat and pressure through the periodic boundaries during deformation.  The research herein explores how the tensile deformation of amorphous polyethylene, modelled with a united atom method potential, is influenced by heat and pressure dissipation.  The stress-strain curves for the pressure dissipation cases (uniaxial tension) are in qualitative agreement with experiments and show that heat dissipation has a large effect on the strain hardening modulus calculated by molecular dynamics simulations.  Moreover, in addition to quantifying the evolution of the energy associated with bonded and non-bonded terms as a function of strain, the evolution of stress associated with these different components in both the loading and non-loading directions was also calculated as a function of strain to give insight into how the stress state is altered within the elastic, yield, strain softening, and strain hardening regions.  The energy partitioning shows that the majority of energy increase during deformation is associated with the non-bonded Van der Waal's interactions, similar to previous studies.  The stress partitioning shows a competition between `tensile' Van der Waal's interactions and `compressive' bond stretching forces, with the characteristic yield stress peak clearly associated with the non-bonded stress.  Subsequent analyses concentrates on the evolution of several internal structure metrics with strain: bond length, bond angle, dihedral conformation, chain orientation, and chain entanglement.  The lack of heat dissipation had the largest effect on the strain hardening regime, where an increase in the calculated temperature correlated with faster chain alignment in the loading direction and more rapid conformation changes.  In part, these observations demonstrate the role that heat and pressure dissipation play on deformation characteristics of amorphous polymers, particularly for the strain hardening regime.

\end{abstract}

\section{Introduction}

Polymers play an important role in materials science as a candidate for future new industrial materials, primarily because their properties can be tuned very easily. The flexibility of tuning polymer's properties comes from a number of available degrees of freedom: choice of monomers, branching, tacticity, copolymers, blends and composites.   This flexibility also means that polymeric systems are highly complex in nature. Computational modeling of polymers has assisted in understanding their complex behavior. However, no single modeling technique, to this point, has adequately captured all of the large length and time scales associated with polymers.  The challenge of modeling polymers in the robust design of industrial components requires a computational strategy that incorporates structure-property relationships over many length scales. For this purpose, a hierarchical multiscale modeling strategy~\cite{Bou2009} that bridges nanoscale mechanisms to the macroscale/continuum by introducing structure-property relationships is very attractive.

Coarse grain techniques~\cite{Uhl1998,Fal2002,Fuk2002,Mul2002}, such as united atom models~\cite{Pau1995,Yoo1993,Li2006,Mul2007,Hos2010}, have been widely used to simulate polymer systems at various length and time scales.  For instance, previous work has used Monte Carlo methods and molecular dynamics (MD) simulations to study deformation mechanisms during uniaxial tensile deformation of amorphous polyethylene-like glassy polymers~\cite{Li2006,Mul2007,Hos2010}. The stress-strain curves qualitatively show agreement with experimental results on amorphous polymers.  Factors such as the energy contributions from the potential, the stress partitioning, and the changes in free volume, chain orientation, and chain entanglements were also calculated as a function of strain to help elucidate the inherent deformation mechanisms.  Insight from these atomistic simulations, such as the nanoscale energy contributions, can be used to inform relevant internal state variables at the continuum level for constitutive models of amorphous glassy polymers~\cite{Bou2010}.  The evolution of these internal state variables aims to capture the evolution of microstructure that occurs as a function of stress state and processing history.

Thus, understanding and quantifying the nanoscale mechanisms associated with plastic deformation in amorphous polymers is important for multiscale models that aim to predict the macroscopic properties of polymers.  Conducting experiments on the molecular level to investigate the deformation mechanisms in polymer systems is often very difficult.  Hence, the use of molecular dynamics (MD) simulations have proven to be essential tools for exploring the static, dynamic and mechanical properties of bulk amorphous polymers at the nanoscale.  Atomistic simulations have provided much needed insight into the plastic deformation behavior of some thermoplastic polymer systems.  For instance, numerous atomistic studies have successfully captured the mechanical properties (Young modulus, yield peak, strain-hardening modulus) of glassy polymers using bead-spring models~\cite{Gao1995,Gao1995a,Gao1996,Rot2001,Rot2003,Hoy2006,Hoy2007,Hoy2008} (with and without the bond angle potential component).  For instance, Hoy and Robbins~\cite{Hoy2006,Hoy2007,Hoy2008} have used such coarse-grained bead-spring models to interrogate strain hardening in glassy polymers as a function of both microstructure and deformation conditions; they found that the plastic flow stress correlated with the strain hardening modulus, indicating that entropic network models used for rubber elasticity theory may not accurately capture the nanoscale physics of strain hardening in polymer glasses.  Additionally, other simulations have used more chemically-realistic potential formulations for studying properties in specific systems (e.g., polyethylene~\cite{Bro1991,Mc1993,Cap2002,Lav2003,Lee2011}) as well as different numerical schemes for molecular deformation, such as Monte Carlo~\cite{Li2006,Mul2007,Chu1999}.  However, although several groups have studied the static, dynamic, and mechanical properties of glassy polymer systems, there is still much that is not understood with respect to polymer deformation simulations.  In multiscale models, it is not only important to pass information from scale-to-scale, but also to understand what effect different assumptions and uncertainties have on the information being passed.

There are a number of potential sources of uncertainty associated with polymer deformation simulations at the nanoscale.  For instance, loading and boundary conditions, polymer characteristics, model parameters, and polymer forcefield may all impact the deformation response in nanoscale simulations.  Much of the literature relating to mechanical properties has focused on the influence of loading conditions and polymer characteristics.  However, not as much literature has focused on how model parameters and polymer forcefields affect the deformation response.  For instance, how does coarse-graining affect properties compared to full-atom representations?  What influence does crystallinity have on the deformation response?  How does a reactive forcefield that includes bond formation and breaking differ in inelastic properties from forcefields that cannot account for this?  There is an increasing number of studies that aim to answer these questions for different polymer systems.  For instance, the influence of interatomic potential parameters and the potential formulation has been studied for a few polymeric material systems~\cite{Yoo1993,Gee1998,Che2006}, but there still remains questions with respect to understanding the influence of coarse-graining and different potential formulations.  Here, we have chosen to investigate one aspect of uncertainty pertaining to how the thermostat and barostat conditions affect the deformation response.  Several studies have focused on uniaxial and triaxial states of stress associated with how pressure is dissipated at the boundary conditions~\cite{Rot2001,Mak2009} or even different deformation protocols~\cite{Mul2007}.  For instance, Rottler and Robbins~\cite{Rot2001} used a bead-spring model with three stress states (uniaxial tension, biaxial compression, and biaxial shear) to examine yield and yield criteria for amorphous polymers.  More recently, Makke et al.~\cite{Mak2009}~used a bead-spring model with both uniaxial and triaxial tensile stress states to examine the effect of boundary-driven deformation.  Additionally, Mulder et al.~\cite{Mul2007} used Monte Carlo simulations to investigate different deformation conditions in a polyethylene-like polymer system: (i) rigid vs.~flexible bonds and (ii) affinely displacing all atoms vs.~displacing the center-of-mass of entire molecules.  However, the role of the thermostat in dissipating heat generated during deformation and how this affects deformation of glassy polymers has not been extensively discussed in the literature.  Moreover, the analysis herein examines the evolution of selected structure metrics (e.g., bond length/angle, conformation, orientation, entanglement) and stress/energy components as a function of strain. 

Therefore, the objective of this research is to explore how the internal structure of amorphous polyethylene is influenced by tensile deformation as well as the ensemble used for the non-loading lateral directions, which introduces overall stress state differences.  While there is a wide range of potential rates for the thermostat and barostat on these lateral directions, we focused on the four extreme ensemble cases that correspond to full/no heat dissipation and full/no pressure dissipation: isothermal-isobaric (NPT) ensemble, isenthalpic-isobaric (NPH) ensemble, canonical (NVT) ensemble, and microcanonical (NVE) ensemble.  Since the loading direction is decoupled from the lateral directions for deformation, the entire simulation cell may not be in the ensemble prescribed for the lateral directions (i.e., the total energy may increase for the isoenergetic ensemble, etc.).  At the strain rates used here, previous simulations by the present authors have shown that the stress-strain behavior with elastic, yield, strain softening and strain hardening regions are qualitatively in agreement with previous simulations and experimental results.  The energy and stress contributions from the united atom potential were calculated as a function of strain to help elucidate the inherent deformation mechanisms within the elastic, yield, and strain hardening regions \textit{and} how these change based on the boundary conditions.  The results of examining the partitioning of stress and energy as well as the changes that accompany the different loading conditions can shed light on the role of ensemble choice on the deformation behavior of amorphous polyethylene.

\section{Simulation Methodology} 
\subsection{Interatomic potential}
The interatomic force field for amorphous PE is based on a united atom model using the DREIDING potential~\cite{May1990}.  There were several reasons why this potential was chosen.  First, this study investigates uniaxial deformation of amorphous PE systems with large numbers of monomers.  Full atom representations would require approximately three times the number of atoms (CH$_2$) along with a more computationally-expensive force field and, quite possibly, a smaller timestep or different integration scheme due to adding hydrogen. Therefore, the use of the united atom model for the forcefield was influenced in part by the computational expense.  Second, prior literature regarding molecular dynamics simulations of PE have used similar potentials~\citep{Li2006,Mul2007,Hos2010}.  In this sense, this work can build upon results previously obtained, while also providing guidance as to the influence of boundary conditions when moving to full atom representations with optimized force fields.

The united atom model used here considers each methyl group (i.e., the CH$_2$ monomer) as a single atom with a force field that determines the interactions between united atoms.  The DREIDING potential ~\cite{May1990} used has four contributing terms; bond stretching, changes in bond angle, changes in dihedral rotation, and van der Waals non-bonded interactions.  The total force field energy can be expressed as
\begin{equation}
	E_{total}=E_{b}(r)+E_{\theta}(\theta)+E_{\phi}(\phi)+E_{nb}(r)
\end{equation}
In the present simulations, the interaction potential has two components which are summed over all possible atom combinations: bonded and non-bonded interaction terms. The bonded terms comprise bond stretching ($r$), bond-angle bending ($\theta$) and dihedral angle torsion ($\phi$). The functional form of bonded energy is given as
\begin{equation}
	E_{b}(r)=\frac{1}{2}K_b\left(r-r_0\right)^2
\end{equation}
\begin{equation}
	E_{\theta}(\theta)=\frac{1}{2}K_\theta\left(\theta-\theta_0\right)^2
\end{equation}
\begin{equation}
	E_{\phi}(\phi)=\sum_{i=0}^3{C_i\left(cos\phi\right)^i}
\end{equation}
where $K_b$ and $K_\theta$ are the stiffness constants for the bond length and bond angle potentials, respectively, $r_0$ and $\theta_0$ are the equilibrium bond length and bond angle, respectively, and the variable $C_i$ contains the coefficients of dihedral multi-harmonic.  For the parameters chosen in the present work, the global minimum of the dihedral potential is at 180$^\circ$, which corresponds to the planar \textit{trans} state. In addition, the two local \textit{gauche} minima are located at 60$^\circ$ and 300$^\circ$.  The non-bonded or Van der Waals interactions are given by a Lennard-Jones potential, i.e., 
\begin{equation}
	E_{nb}(r_{ij})=4\epsilon_0\left[\left(\frac{\sigma_0}{r_{ij}}\right)^{12}-\left(\frac{\sigma_0}{r_{ij}}\right)^{6}\right], r_{ij}\leq{r_c}
\end{equation}
where $r_{ij}$ is the distance between atoms $i$ and $j$, $\sigma_0$ is the zero energy spacing of the potential, and $\epsilon_0$ is the energy well depth of the potential. The cutoff distance $r_c$ is taken as 10.5 \AA.  The parameters for the PE force field are given in Table \ref{tab:InteratomicPotentialParameters}.  

\subsection{Deformation simulations}

A parallel molecular dynamics code, LAMMPS ~\cite{Pli1995}, was used to deform the polymer simulation cells.  The simulations were performed for a three-dimensional periodic simulation cell with 100 polyethylene chains of 1000 monomers each, for a total of $10^5$ united atoms.  The temperature chosen for this study was 100 K.  This temperature is sufficiently below the calculated glass transition temperature ($T_g=300$ K~\cite{Hos2010}).  Therefore, the amorphous polyethylene response is representative of that in the glassy state.  The initial chain structure was created using a method similar to those previously developed using Monte Carlo self-avoiding random walks ~\cite{Bin1995}.  The chain generation step used a face-centered cubic (FCC) lattice superimposed on the simulation cell with the nearest neighbor distance of $1.53$ Angstroms.  Molecules were added to the lattice in a probabilistic stepwise manner [e.g., Ref.~\citenum{The1985}] that based the probability of chain growth in certain directions on the bond angle and the density of unoccupied sites in the region.  

The initial polymeric structure was then inserted into the molecular dynamics code where an equilibration sequence was performed prior to deforming the amorphous polymer.  The equilibration sequence relaxes any high energy configurations that are artificially created due to the face-centered cubic lattice used to generate the amorphous polymer structure.  The relaxation involves four different steps. Initially, the simulation ran for $10^5$ timesteps ($\Delta{t}=1$ fs) using NVT dynamics at 500 K followed by relaxation for $5$x$10^5$ timesteps ($\Delta{t}=0.5$ fs) using NPT dynamics at 500 K. The next relaxation cooled the structure down to the desired temperature for $5$x$10^5$ timesteps followed by further relaxation of $5$x$10^5$ timesteps ($\Delta{t}=1$ fs) at 100 K.
  
Several microstructure metrics were used to quantitatively validate the polymer chain geometry following equilibration.  The average bond length for all equilibrated systems was 1.529 \AA\ and the average bond angle was 109.27$^\circ$.  The average values were slightly lower than the equilibrium potential parameters $r_0$ and $\theta_0$.  The dihedral angle distribution has a broad distribution with both \textit{gauche} and \textit{trans} peaks.  To calculate the fraction of \textit{trans} conformations within the PE system, a threshold value of 120$^\circ$ was used to delineate the \textit{gauche} peak (centered about 66$^\circ$) from the \textit{trans} peak (centered about 180$^\circ$).  The percent \textit{trans} conformations in the initial structure was 73.8\%.  The initial densities for the amorphous PE structures ranged from 0.87-0.91 g/cm$^3$, slightly lower than experimental values for low density PE (0.91-0.94 g/cm$^3$), which contains a high degree of short and long chain branching.  The glass transition temperature was used to verify that such equilibration of the polymer structure was appropriate for deformation simulations.   The glass transition temperature $T_g$, delimiting the glassy regime to the rubbery one, was determined from the change in slope of the specific volume versus temperature curve.  The calculated values of $\approx300$ K here are similar to those calculated by other groups using molecular dynamics, which fall in the range of 250-300 K~\cite{Lav2003,Gee1998,Tak1991,Han1994}, close to the experimentally-measured value of 250 K~\cite{Bra1989}.

\begin{figure*}[t]
  \centering
\includegraphics[width=0.8\columnwidth,angle=0]{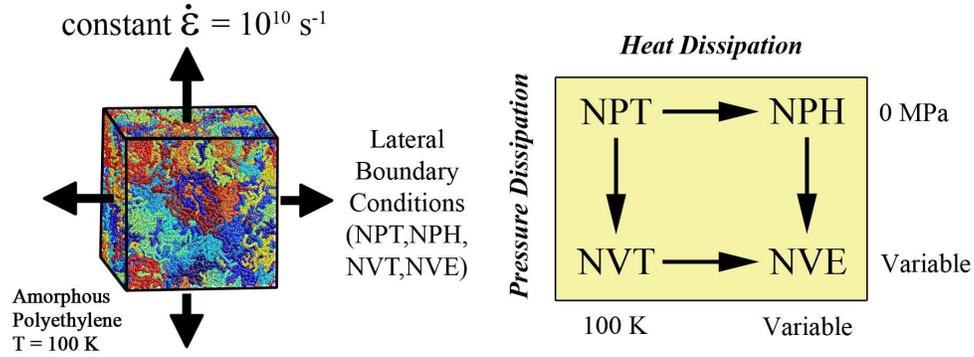}
\caption{Schematic of simulation cell and boundary conditions.  The simulation cell is loaded at a constant strain rate in one direction and the lateral boundaries are controlled by four ensembles representing the extreme cases of heat and pressure dissipation.}
  \label{fig:fig1}
\end{figure*}

The amorphous PE system was then deformed under tensile loading applied at a constant true strain rate with different boundary conditions for the two lateral simulation cell faces, as shown schematically in Fig.~\ref{fig:fig1}.  The NPT deformation condition allows for both rapid pressure and heat dissipation (in comparison to the loading) leading to zero pressure on the lateral boundaries and a constant 100 K within the simulation cell.  Essentially, this is a pure uniaxial tensile simulation with rapid heat dissipation within the sample (plane stress condition).  In contrast to NPT, the NPH boundary condition does not regulate the system temperature, allowing the heat generated through internal work to increase within the sample.  Again, this boundary condition is a pure uniaxial tensile simulation where heat does not have adequate time to diffuse.  In reality, the rate of heat dissipation falls somewhere in between; these conditions were chosen to bound this behavior.  Moreover, one must realize that multiple cases may occur within the same material due to constraints caused by various microstructure heterogeneities.

The NVT and NVE boundary conditions have no pressure dissipation on the lateral boundaries.  Thus, these boundary conditions impose a triaxial stress state as the stress increases on the constrained lateral boundaries as a function of strain (plane strain condition).  Similar to the NPT and NPH boundary conditions, the NVT and NVE boundary conditions have rapid heat dissipation and no heat dissipation, respectively.  For the NPT and NVT conditions, a Nos$\acute{e}$-Hoover thermostat was used to regulate the system temperature~\cite{Nos1984,Hoo1985}.  For the NPT and NPH boundary conditions, the pressure dissipation was implemented by decoupling the boundary in the loading direction from the equations of motion from the other two directions~\cite{Mel1993}, similar to that used in previous simulations~\cite{Yan1997,Mak2009}. 

Several stress and energy measures were tracked to examine their dependence on the deformation boundary condition used.  The macroscopic stress components were calculated from the symmetric pressure tensor, which uses components from the kinetic energy tensor and the virial tensor.  The total energy, kinetic energy, and potential energy were also recorded.  In addition to calculating these macroscopic measures, the contributions to the total stress and energy from the bond lengths, bond angles, dihedral angles and non-bonding interactions were also tracked as a function of strain.  More details on how this partitioning is performed for the macroscopic stress tensor are given elsewhere~\cite{Tho2009}. Periodic dumps of the atomic configuration were performed to compute additional microstructure metrics such as chain orientation or the percentage of \textit{trans} dihedral conformations.  An effective true strain rate of $10^{10}$ s$^{-1}$ was used to deform the amorphous polyethylene simulation cells.  To reduce the variability observed in the stress and energy responses, multiple steps were taken here: 
\begin{enumerate}
	\item Four different initial configurations were used to create multiple instantiations of amorphous polyethylene
	\item Each configuration was separately deformed in the x, y, z directions to generate a total of twelve different PE deformation datasets
	\item The strain in the loading direction was applied every 10 timesteps to allow sufficient time for the lateral boundaries to relax
	\item Fluctuations in the stress and energy calculations were reduced using a local regression technique with a 2$^{nd}$ degree polynomial model
	\item Stress and energy responses for each condition resulted from averaging the filtered responses for the twelve different PE deformation datasets 
\end{enumerate}

\begin{figure}[!t]
  \centering
\includegraphics[width=3.25in,angle=0]{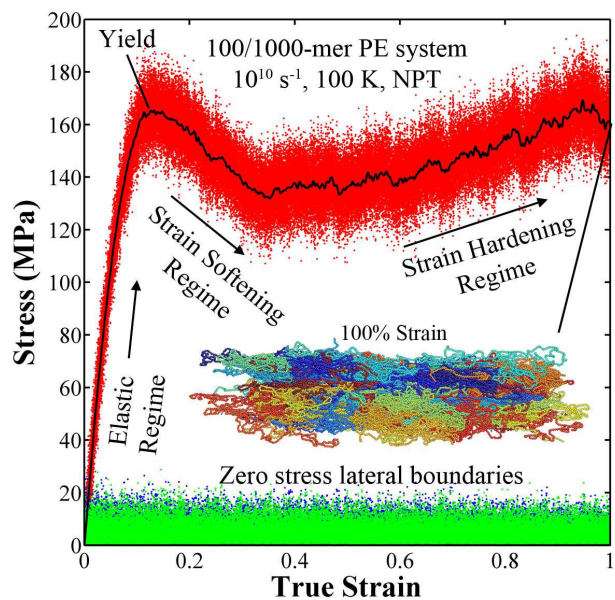}
\caption{Stress-strain response of amorphous polyethylene deformed in uniaxial tension (NPT condition).  The accompanying image shows the polyethylene structure at 100\% strain (colors represent united atoms on separate chains).}
  \label{fig:fig1a}
\end{figure}

The stress-strain response for a 100-chain, 1000-monomer per chain amorphous PE system is shown in Fig.~\ref{fig:fig1a}.  This system was deformed at 100 K and 10$^{10}$ s$^{-1}$ strain rate using NPT boundary conditions on the lateral boundaries.  Both the original stress values at each strain and the post-processed stress values are shown in this plot to show the reduction in stress fluctuations used here.  Afterwards (not shown in Fig.~\ref{fig:fig1a}), the stress-strain curves for the twelve different PE deformation datasets were averaged to obtain the overall stress behavior.  The stress-strain curve has four distinct regimes typical of experimental curves: elastic, yield, softening and hardening.  Initially, in the elastic regime, the stress increases nearly linearly with increasing applied strain.  Interestingly, the characteristic yield peak observed in experiments is also observed at strain rates of 10$^{10}$ s$^{-1}$, but the size of this yield peak decreases with decreasing strain rate, as detailed in previous studies~\cite{Hos2010}.  Moreover, the stress-strain curves obtained are similar to those obtained in other studies using coarse-grained models of amorphous polymers.  Upon reaching the yield point, the stress then shows a decrease in stress for the strain softening regime.  Further deformation of PE causes an increase in stress during the strain hardening regime.  Also shown in this plot is the stress on the lateral boundaries, which also fluctuates around zero stress for the NPT and NPH conditions.  The described methodology for reducing variability was applied to all subsequent boundary conditions in this paper.

\section{Simulation Results} 

\subsection{Stress-strain and temperature response}

The overall stress-strain behavior and temperature evolution for a 100-chain, 1000-monomer per chain amorphous PE system with different lateral boundary conditions is shown in Fig.~\ref{fig:fig3}.  First, we examine the deformation simulations where the pressure was dissipated through the lateral boundaries (NPT/NPH).  For the NPT boundary condition, notice that all four regimes are observed for the 100 chain, 1000-monomer PE system.  For the NPH condition, the stress-strain behavior is very similar in the elastic, yield, and strain softening regimes, but the stress does not increase in the strain hardening regime in contrast to the NPT case.  Interestingly, comparing the temperature evolution between the two conditions with respect to strain, we observe that the temperature increases drastically for the NPH boundary condition, inducing material softening at large strains.  Next, we examine the deformation simulations where there is no pressure dissipated at the lateral boundaries (NVT/NVE).  For both conditions, the triaxial stress-strain response exhibits a nonlinear elastic response up to a peak stress, with a rapid decrease in stress thereafter.  The stresses in both the tensile direction and the non-loading directions are high ($\approx200$ and $\approx250$ MPa, respectively) due to the lack of pressure dissipation in the lateral boundaries.  This induces a high triaxial state of stress in the deformed PE system.  Additionally, there is little difference in temperature evolution prior to the peak stress for these boundary conditions.  Moreover, the temperature difference observed for larger strains does not significantly impact the overall stress-strain response for the NVT/NVE conditions.   

Fig.~\ref{fig:fig10} shows images of the 100-chain, 1000-monomer per chain amorphous PE system as a function of strain for the NPT and NVT cases.  These two cases were selected to show the differences between the two stress-strain curves.  The conditions without temperature dissipation (NPH, NVE) were qualitatively similar.  In these images, the PE chains were unwrapped through the periodic boundaries and the visualization software AtomEye~\cite{Li2003} was used to render the images.  There are distinct differences between the two boundary conditions, as expected.  The NPT boundary condition allows the pressure to dissipate in the lateral directions as the PE is stretched in the tensile direction.  On the other hand, the NVT boundary condition fixes the bounds in the lateral directions, which imparts a high triaxial state of stress inducing void nucleation and growth, and stretching of the chains.  Notice that the 25\% strain level occurs after the peak stress for the NVT boundary condition.  Prior to the peak stress, the PE system deforms in a relatively uniform manner.  However, upon reaching the peak stress, the PE system deforms unevenly, with the volume associated with the chain stretching accommodating much of the deformation.

\subsection{Internal energy partitioning}

\begin{figure}[t]
  \centering
		\begin{tabular}{c}
		\includegraphics[width=3in,angle=0]{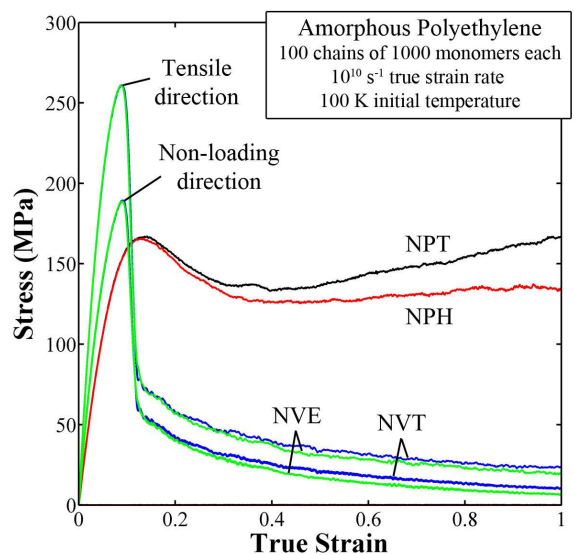} \\
		(a) \\
		\includegraphics[width=3in,angle=0]{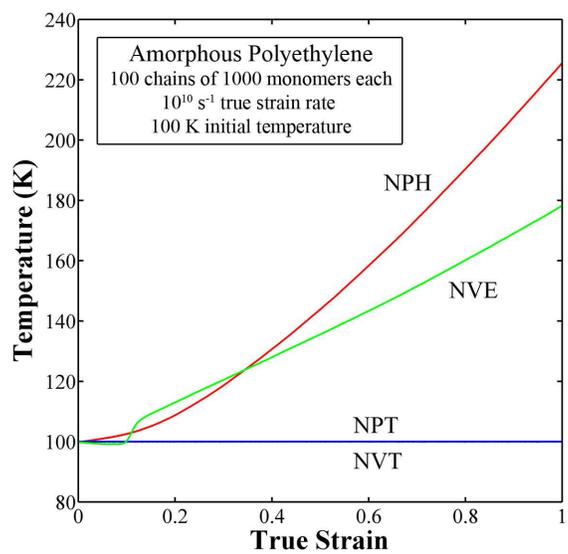} \\
		(b)
		\end{tabular}
\caption{(a) Stress and (b) temperature evolution as a function of true strain for polyethylene.  All four boundary conditions (NPT, NPH, NVT, NVE) are shown and labeled.}
  \label{fig:fig3}
\end{figure}

\begin{figure*}[t]
  \centering
	\includegraphics[width=4in,angle=0]{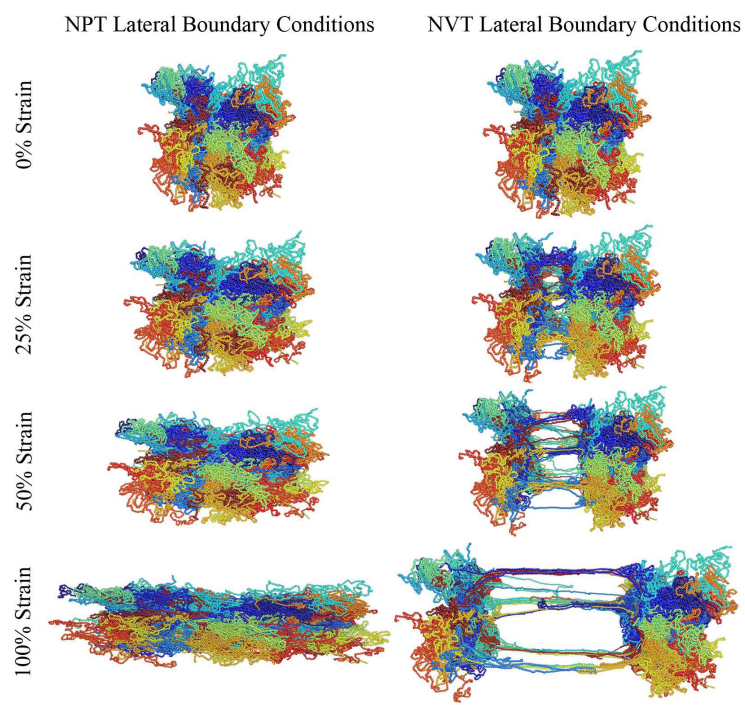}
	\caption{Polyethylene structure evolution for the NPT and NVT lateral boundary conditions at 0\%, 25\%, 50\%, and 100\% strain.}
  \label{fig:fig10}
\end{figure*}

\begin{figure*}[t]
  \centering
		\begin{tabular}{cc}
		\includegraphics[width=3in,angle=0]{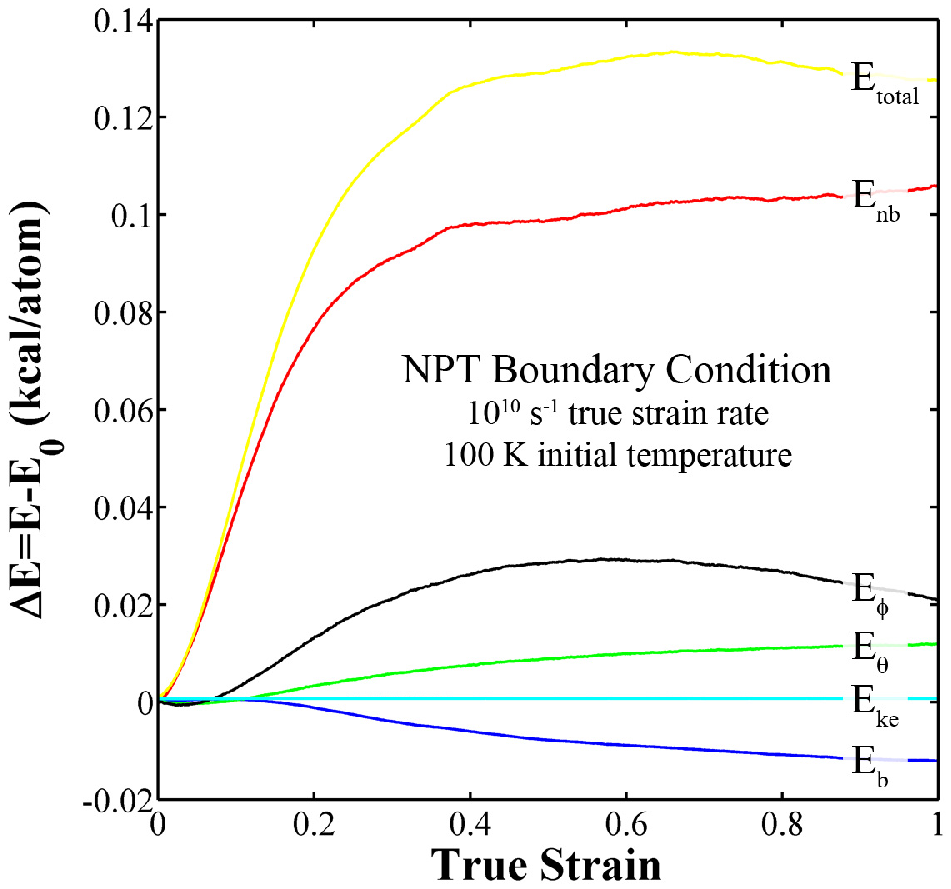} & \includegraphics[width=3in,angle=0]{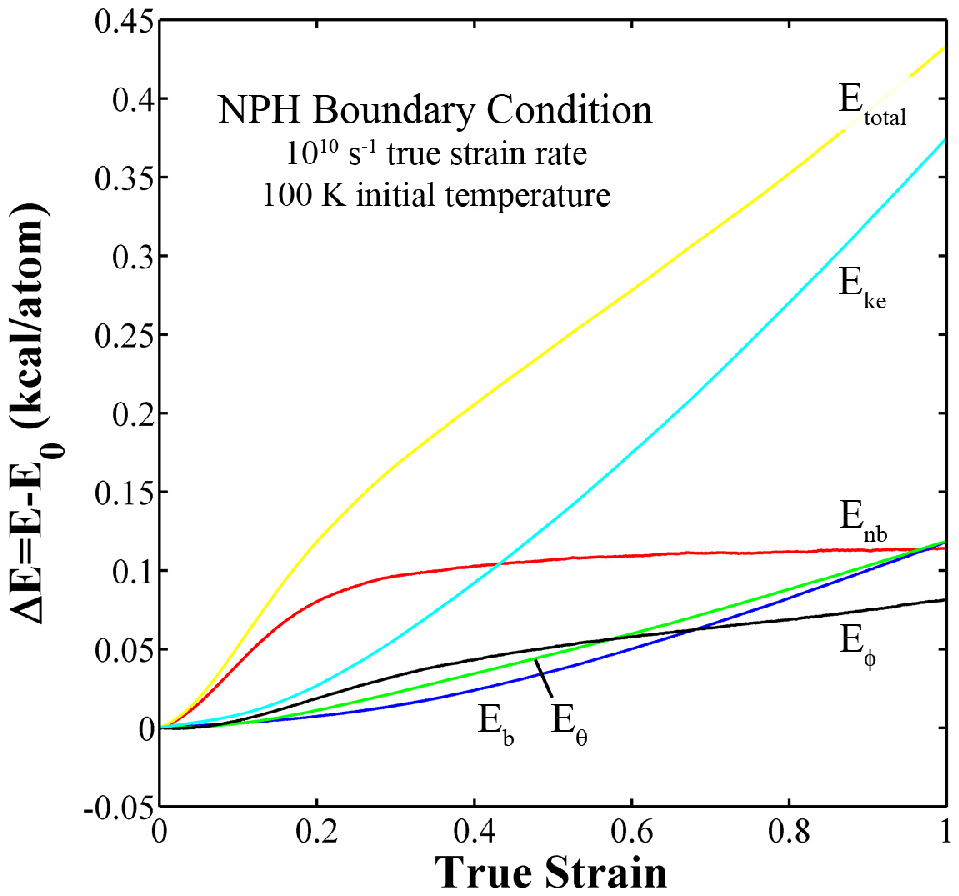}\\
		(a) & (b) \\
		\includegraphics[width=3in,angle=0]{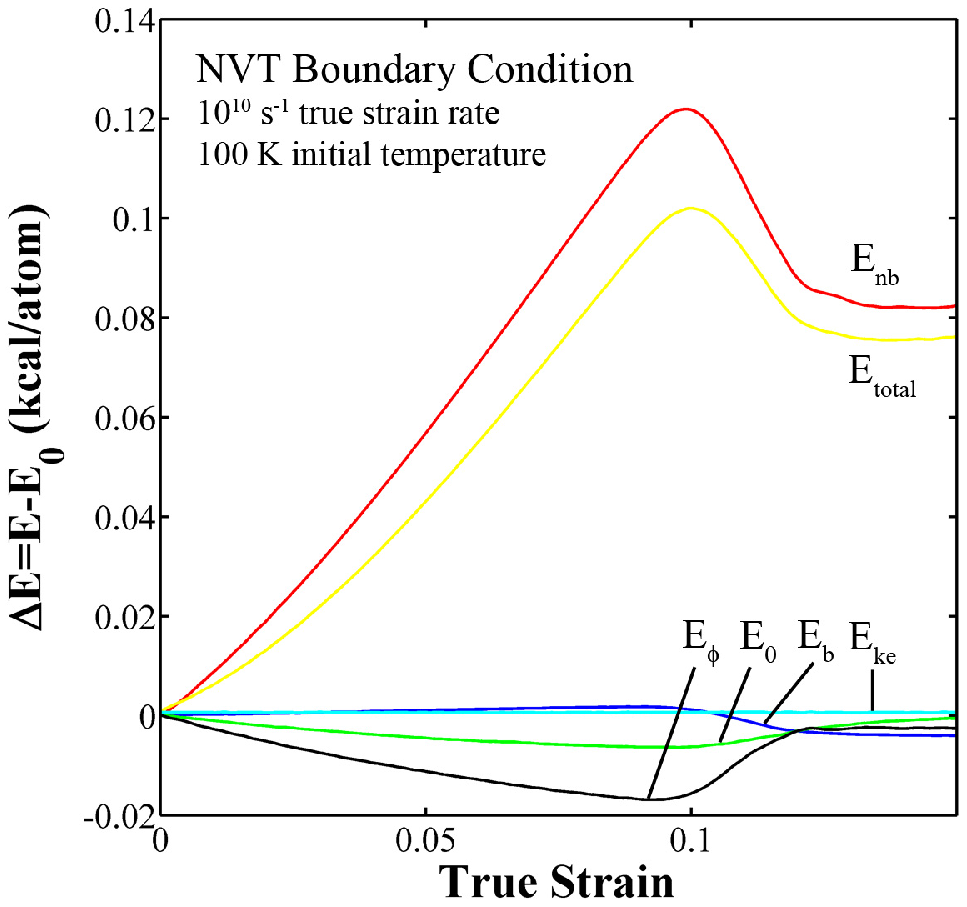} & \includegraphics[width=3in,angle=0]{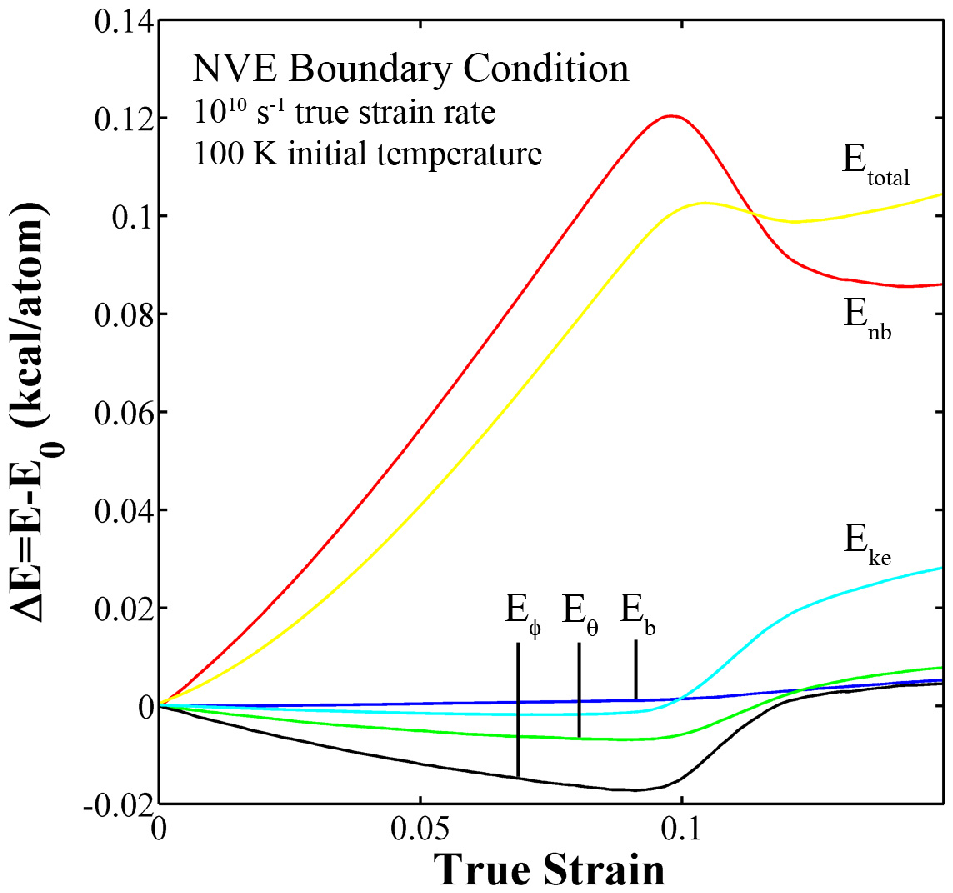} \\
		(c) & (d)
		\end{tabular}
\caption{Normalized energy evolution as a function of true strain for polyethylene for the (a) NPT, (b) NPH, (c) NVT, and (d) NVE conditions.}
  \label{fig:fig3a}
\end{figure*}

Fig.~\ref{fig:fig3a} shows the normalized potential energy change of a 100-chain, 1000-monomer per chain amorphous PE system deformed for a strain rate of 10$^{10}$ s$^{-1}$ at 100 K, which corresponds to the stress-strain behavior observed in Fig.~\ref{fig:fig3}.  In addition to the total energy $E_{total}$, the individual components of the energy are shown: the non-bonded energy $E_{nb}$, the bonded energy $E_b$, the bond angle energy $E_\theta$, the dihedral energy $E_\phi$, and the kinetic energy $E_{ke}$.  The normalized energy, $\delta{E}$, in these plots is obtained by subtracting the initial energy $E_0$ of each component at zero strain.  Therefore, each component's increase/decrease is relative to its initial energy.  The NPT boundary condition in Fig.~\ref{fig:fig3a} will be discussed first.  In the elastic and yield regimes, the dihedral, angle and bond energies remain almost unchanged and a majority of the increase in total energy is correlated with an increase in the non-bonded energy associated with Van der Waal's forces between polymer chains.  The non-bonded energy increases sharply in the elastic and strain softening regions, which can be associated with chain slippage mechanisms.  After the strain softening region, the non-bonded energy stays fairly constant while there are significant changes in the bond length, bond angle and dihedral energies.  Beyond the elastic regime, the energy associated with bond length steadily decreases as bond lengths move towards the equilibrium bond length.  The dihedral angle energy increases in the strain softening regime and then steadily decreases in the strain hardening regime as \textit{gauche} conformations rotate to the lower energy \textit{trans} conformations.  The total energy sharply increases in the elastic region due to non-bonded interactions and then slowly decreases over the strain softening and hardening regions as energy is dissipated via dihedral rotations and bond length changes to accommodate deformation.  

The influence of heat dissipation of the lateral boundaries is evidenced by comparing the NPT and NPH conditions (Fig.~\ref{fig:fig3a}(a) vs.~Fig.~\ref{fig:fig3a}(b)) as well as the NVT and NVE conditions (Fig.~\ref{fig:fig3a}(c) vs.~Fig.~\ref{fig:fig3a}(d)).  As can be expected, the lack of heat dissipation leads to an increase in the internal energy of the system as deformation proceeds, in particular to the kinetic energy component, $E_{ke}$.  The NVT and NVE cases are very similar in their energy evolution response for strains up to 10\%, which is as expected considering that the temperature does not change significantly (Fig.~\ref{fig:fig3}(b)).  After these strains, the kinetic energy component steadily increases for the NVE case.  The NPT and NPH cases shows a similar response during the elastic regime, with non-bonded Van der Waal's energy making up a significant portion of the total energy of the system.  However, with increasing strain and increasing kinetic energy, there is also a large increase in the energies associated with the bond length, bond angle and dihedral angle components.  Since the temperature is increasing for the material, the corresponding entropy associated with the bond length, bond angle and dihedral angle constraints results in higher energies for these components.  

The influence of pressure dissipation of the lateral boundaries is evidenced by comparing the NPT with the NVT conditions (Fig.~\ref{fig:fig3a}(a) vs.~Fig.~\ref{fig:fig3a}(c)).  The response is initially similar with the non-bonded Van der Waal's interactions making up a significant portion of the total energy of the system.  All other energy components for the NVT case either remain constant or decrease in energy up to 10\% strain.  This is a sign that the increasing free volume within the system allows for some relaxation of bond and dihedral angles, which in turns lowers the energy of these components.  Interestingly, a significant amount of internal energy (heat) is not generated during the elastic portion of the stress-strain curve (as denoted by the lack of increase in the kinetic energy), but rather heat is only generated as a consequence of mechanisms associated with plastic deformation.

\subsection{Stress partitioning evolution}

The different stress components that make up the macroscopic stress can also be tracked as a function of strain to provide insight into the stress accommodation during deformation.  This partitioning is based on previous studies that have examined the evolution of interchain, intrachain, and kinetic stress components~\cite{Li2006,Chu1999,Vor2009}.   Here, we tracked five terms used in the calculation of the macroscopic stress tensor: bond stretching, bond angle, dihedral rotation, van der Waals non-bonded interactions, and the kinetic component.

\begin{figure*}[t]
  \centering
		\begin{tabular}{cc}
		\includegraphics[width=3in,angle=0]{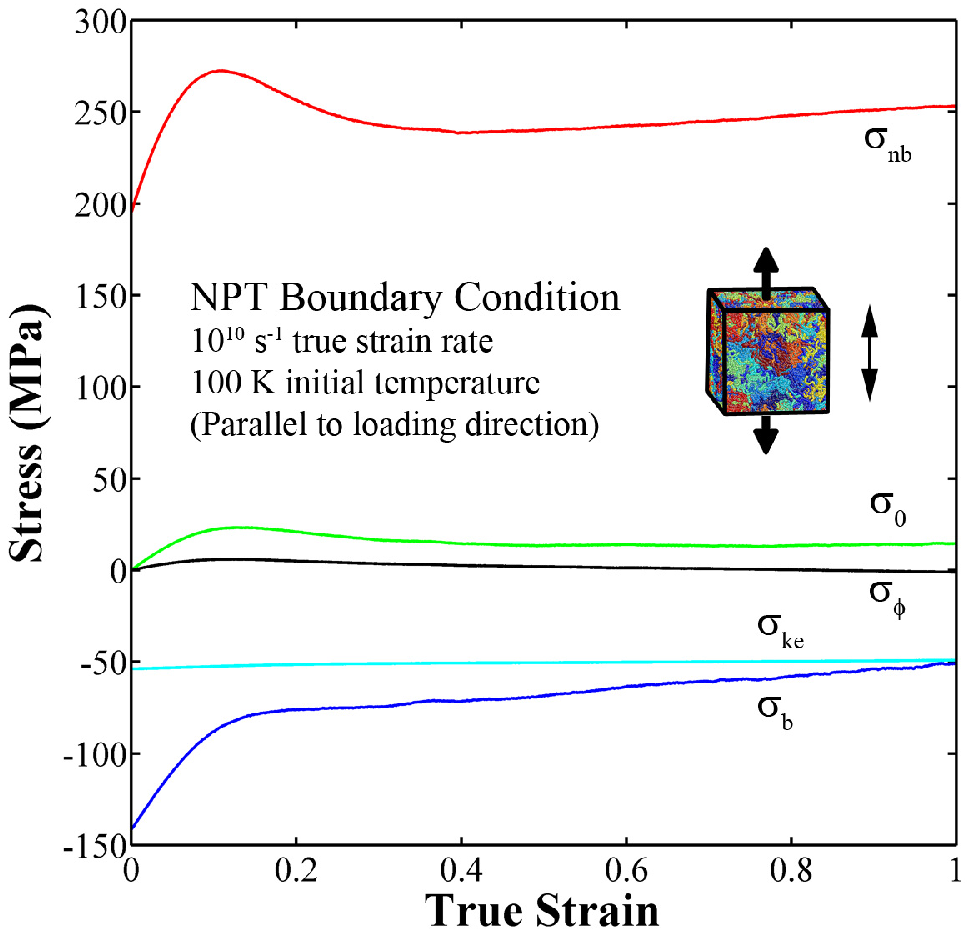} & \includegraphics[width=3in,angle=0]{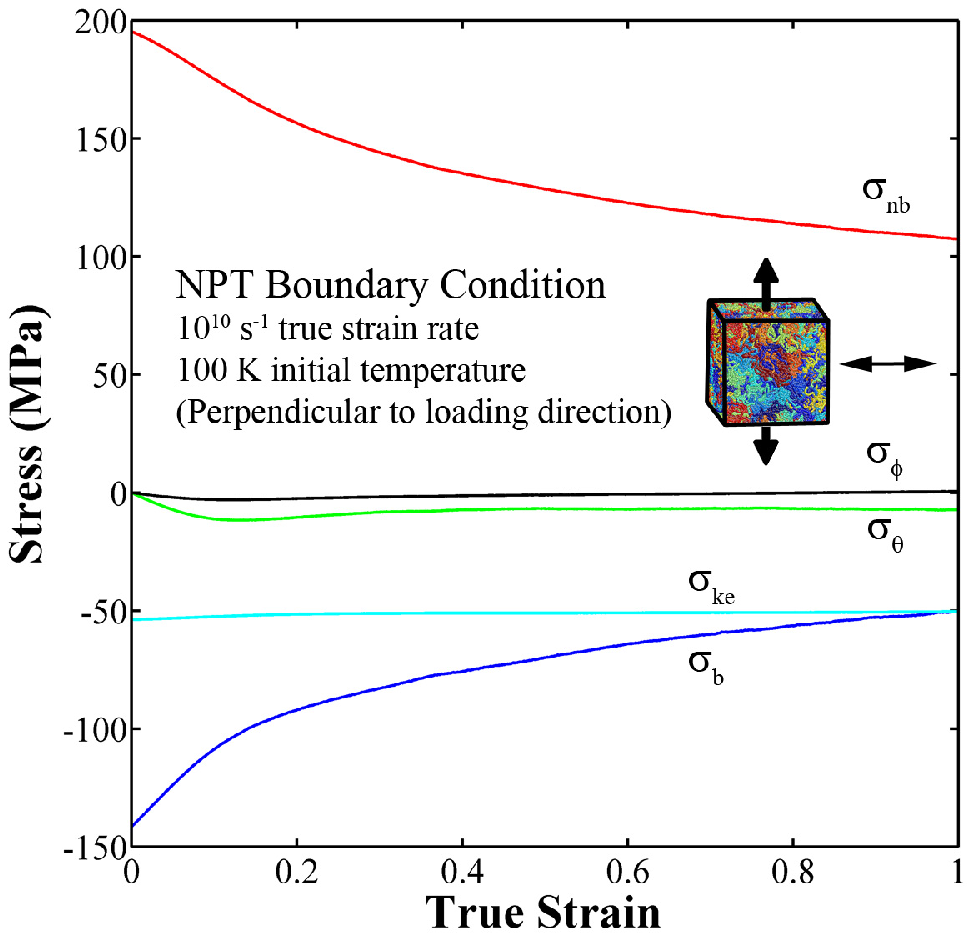}\\
		(a) & (b) \\
		\end{tabular}
\caption{Stress evolution in the (a) tensile direction and (b) lateral directions as a function of true strain for polyethylene.  The macroscopic stress is largely influenced by the `tensile' non-bonded Van der Waal's interactions and the `compressive' bond stretching term.  The characteristic yield peak is associated with the Van der Waal's interactions.}
  \label{fig:fig6}
\end{figure*}

Fig.~\ref{fig:fig6} shows the stress component evolution as a function of strain for the NPT boundary condition.  The two plots show the evolution of the stress components (a) parallel in the direction of loading and (b) perpendicular to the direction of loading.  For the values perpendicular to the loading direction, values from both perpendicular axes were used for the remainder of the paper, i.e., 24 values were averaged instead of 12.  While the response shows a similar trend for several of the stress components ($\sigma_{b}$, $\sigma_{ke}$, $\sigma_{\theta}$, and $\sigma_{\phi}$), the evolution of the stress component attributed to the Van der Waal's interactions ($\sigma_{nb}$) is different between parallel and perpendicular directions.  Notice that for determining the energy relationships, the energy components are non-directional.  Clearly, this plot shows that certain stress components are anisotropic - in particular, the stress components associated with the Van der Waal's interactions and bond angles change as a function of direction.  The fact that these stress components change with direction means that the relationship between energy and stress is not straightforward.  For instance, Fig.~\ref{fig:fig3}(a) shows that the non-bonded energy increases with increasing strain, but this results in both an increase and decrease in the stress associated with this component in directions parallel and perpendicular to loading, respectively.

Interestingly, the bond length stress component $\sigma_{b}$ is the second largest stress contributing the system stress with a large compressive value associated with it.  Moreover, the bond length stress component changes significantly as a function of strain.  The bond angle and dihedral angle stress components ($\sigma_{\theta}$ and $\sigma_{\phi}$) have very small values comparatively and change very little with increasing strain.  In general, the stress partitioning evolution portrays a very different picture from the normalized energy evolution curves; one in which the system stress is largely controlled by the competition between the bond stretching term and the non-bonded Van der Waal's interaction term.  That is, the Van der Waal's forces act to push the polymer chains apart leading to a \textit{tensile} stress, while the bond stretching forces act to keep the chains together leading to a \textit{compressive} stress.

\begin{figure*}[t]
  \centering
		\begin{tabular}{cc}
		\includegraphics[width=3in,angle=0]{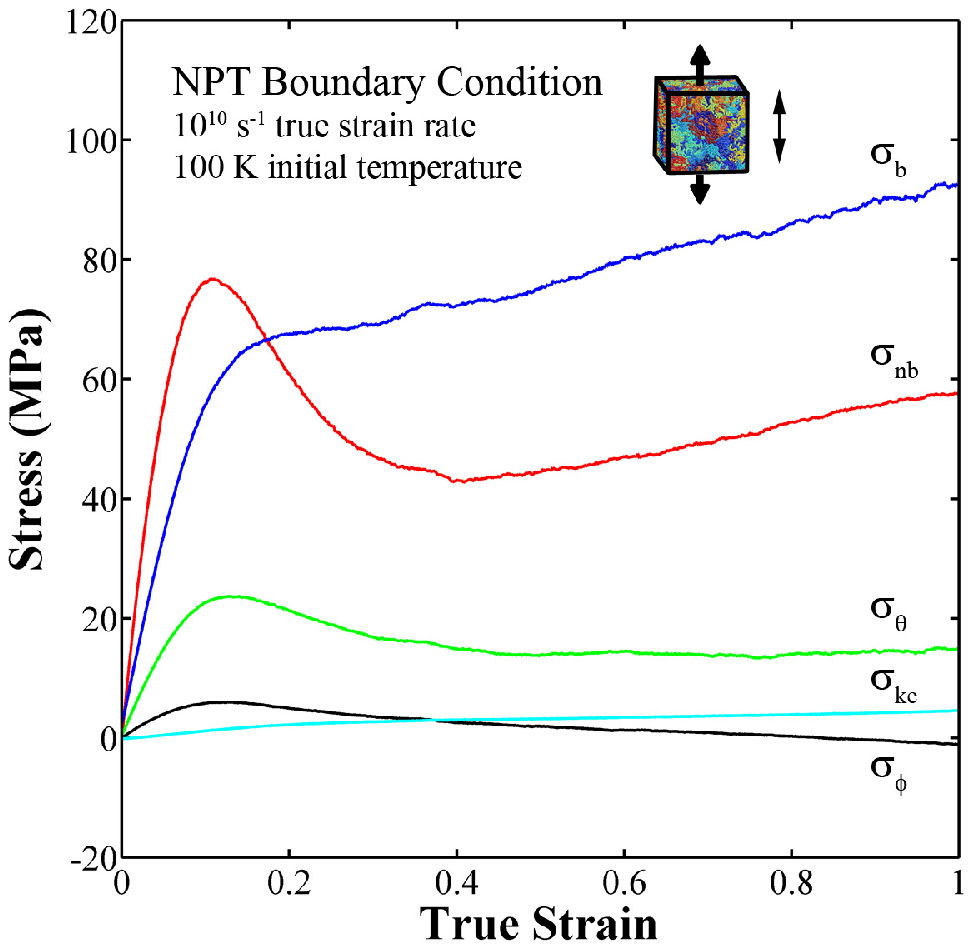} & \includegraphics[width=3in,angle=0]{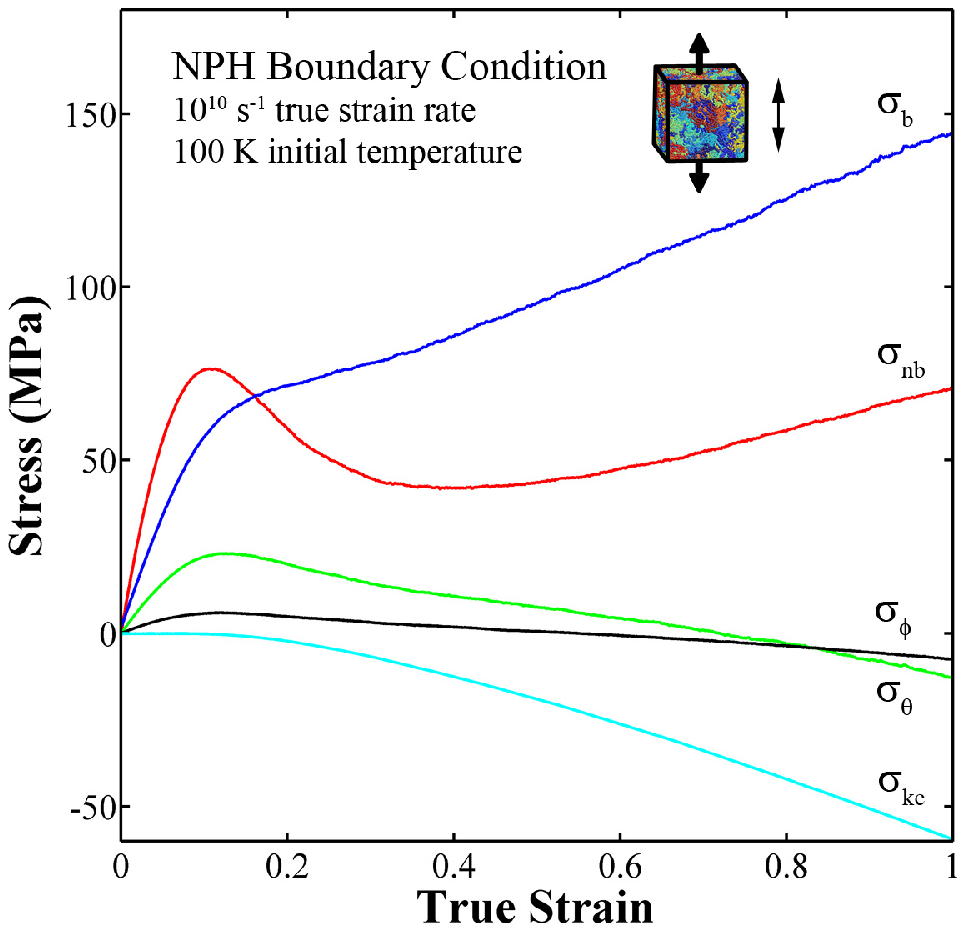}\\
		(a) & (b) \\
		\includegraphics[width=3in,angle=0]{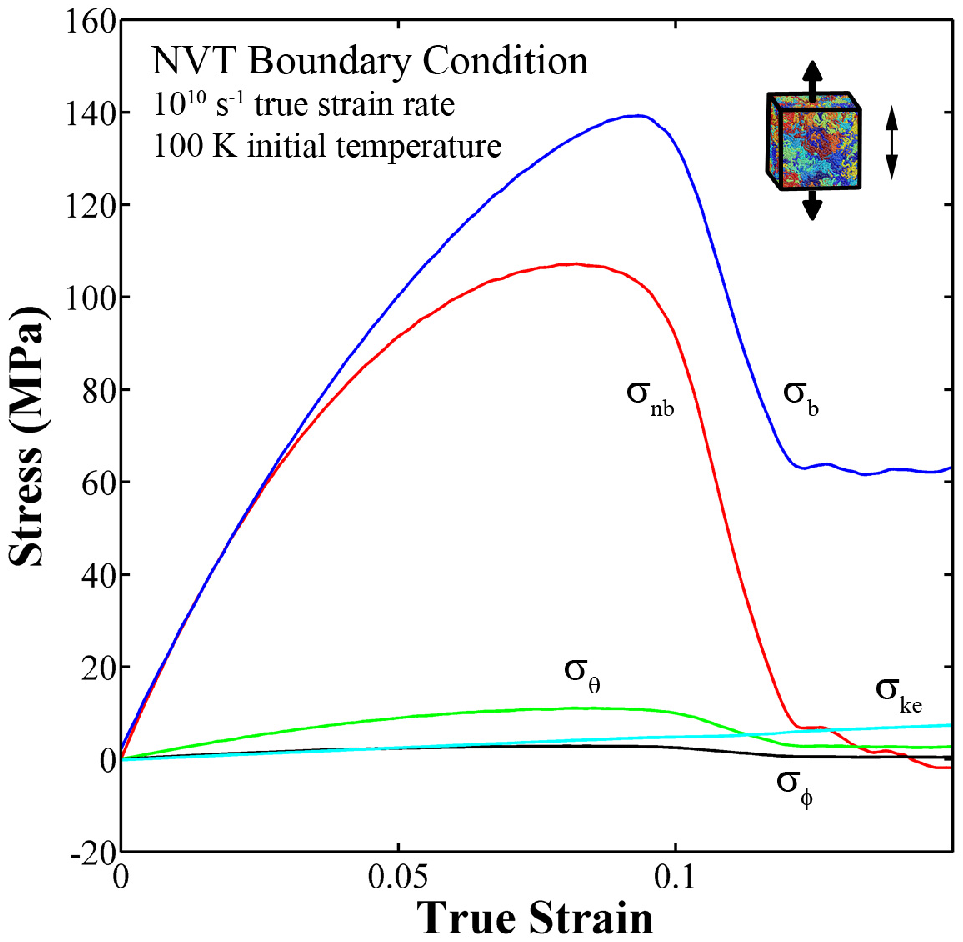} & \includegraphics[width=3in,angle=0]{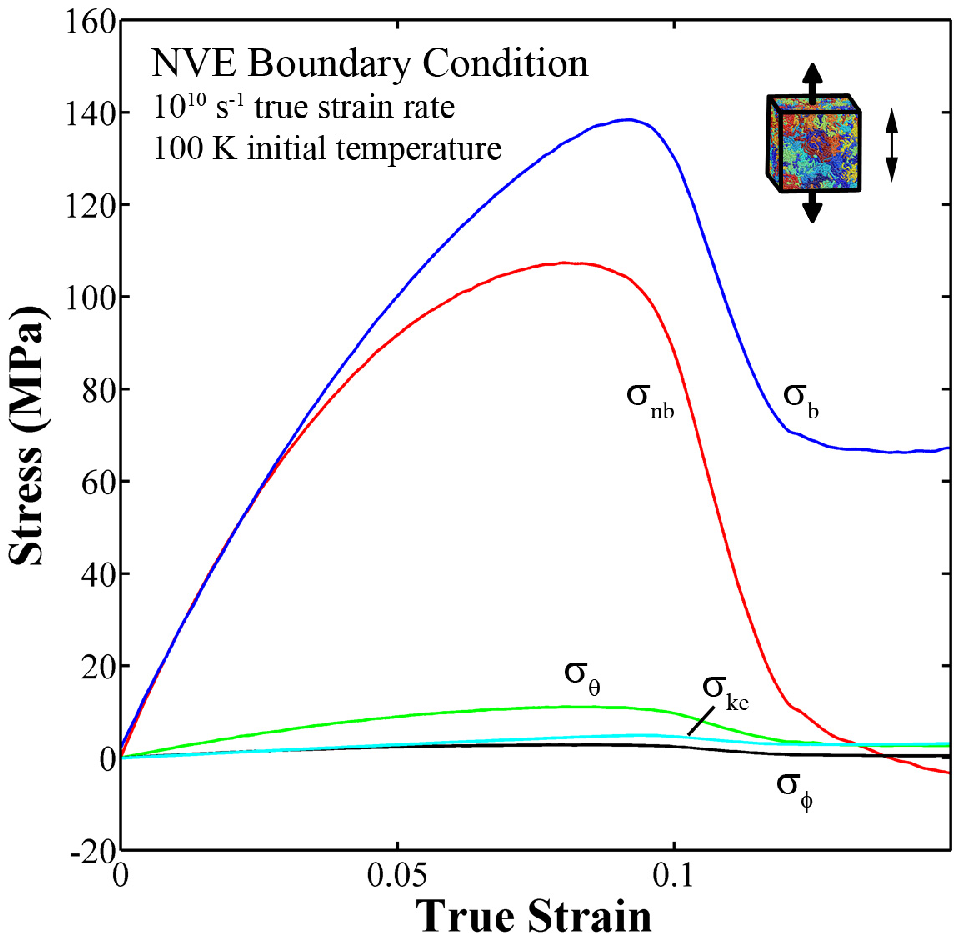} \\
		(c) & (d)
		\end{tabular}
\caption{Normalized stress evolution in the tensile direction as a function of true strain for polyethylene for the (a) NPT, (b) NPH, (c) NVT, and (d) NVE conditions. These curves have been normalized by subtracting the stress values for each component at $\epsilon=0$.}
  \label{fig:fig3b}
\end{figure*}

Fig.~\ref{fig:fig3b} shows the stress component evolution as a function of strain in the direction parallel to the loading direction for all boundary conditions.  These values have been normalized by subtracting their corresponding stress value at zero stress.  First, for all cases, the bond stretching, bond angle, and non-bonded stress components increase with increasing strain up to approximately 10\% strain, with the bond stretching and non-bonded terms being the most significant.  Interestingly, the characteristic strain softening regime of polymers is attributed to the change in the non-bonded stress component, as it then decreases with increasing strain until about 40\% strain and then continues to increase (Fig.~\ref{fig:fig3b}(a) and (b)).  This increase may be correlated to chain crystallization at large strains.  The bond stretching stress component increases monotonically with increasing strain; however, the slope of the stress-strain relation decreases following the yield peak.  For the NVT and NVE simulations, there is little difference for strains up to 10\%.  For the NPH boundary condition, the kinetic energy stress component decreases with increasing strain, which offsets the increases to the bond stretching term (due to a faster rate of chain alignment, shown later).  

\begin{figure*}[t]
  \centering
		\begin{tabular}{cc}
		\includegraphics[width=3in,angle=0]{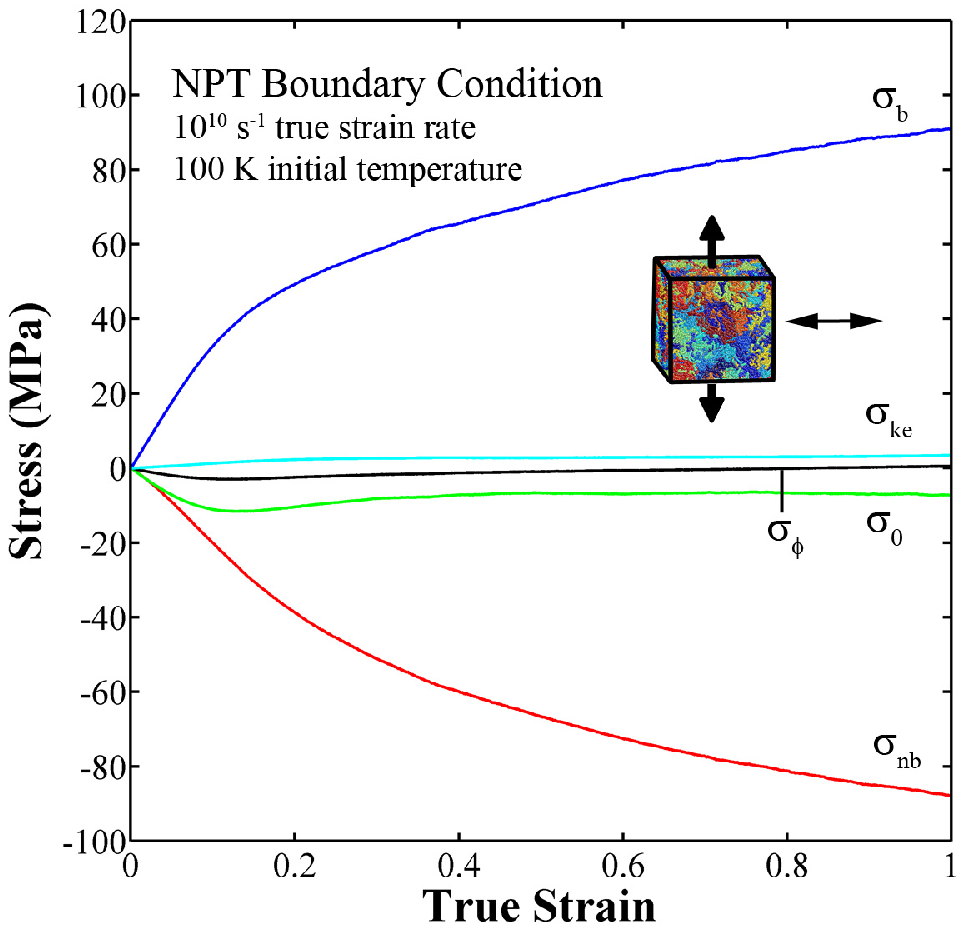} & \includegraphics[width=3in,angle=0]{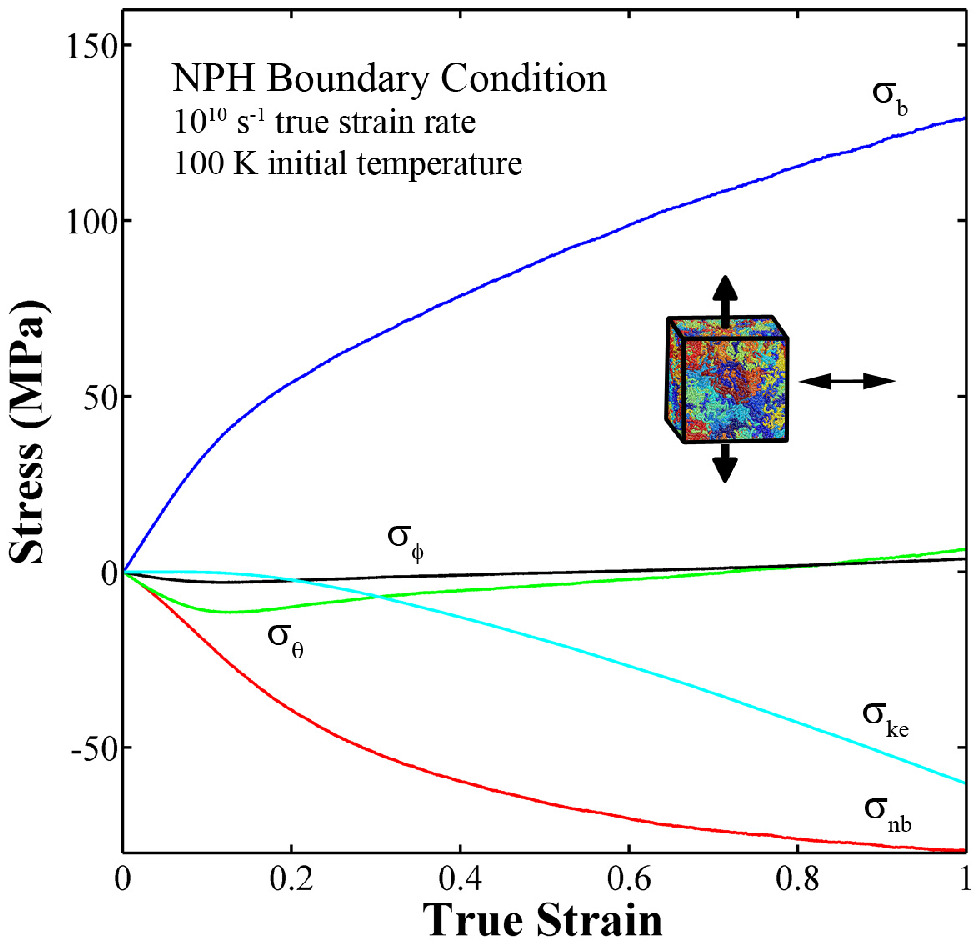}\\
		(a) & (b) \\
		\includegraphics[width=3in,angle=0]{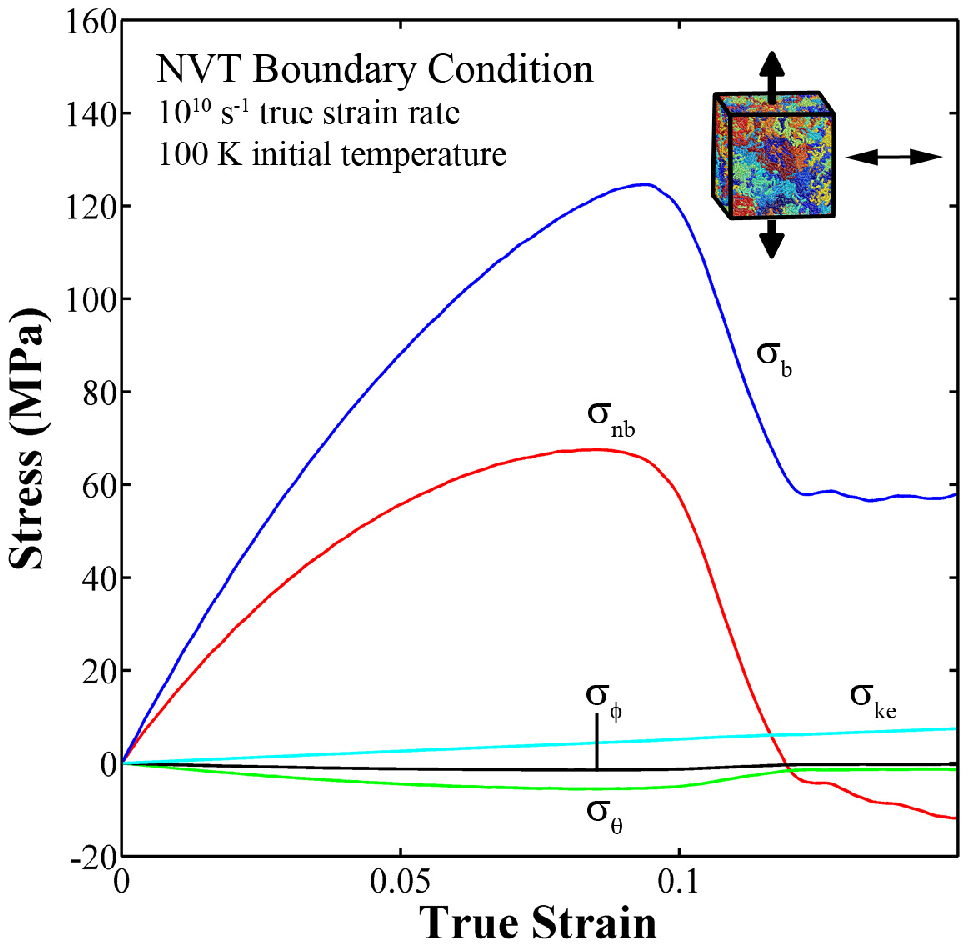} & \includegraphics[width=3in,angle=0]{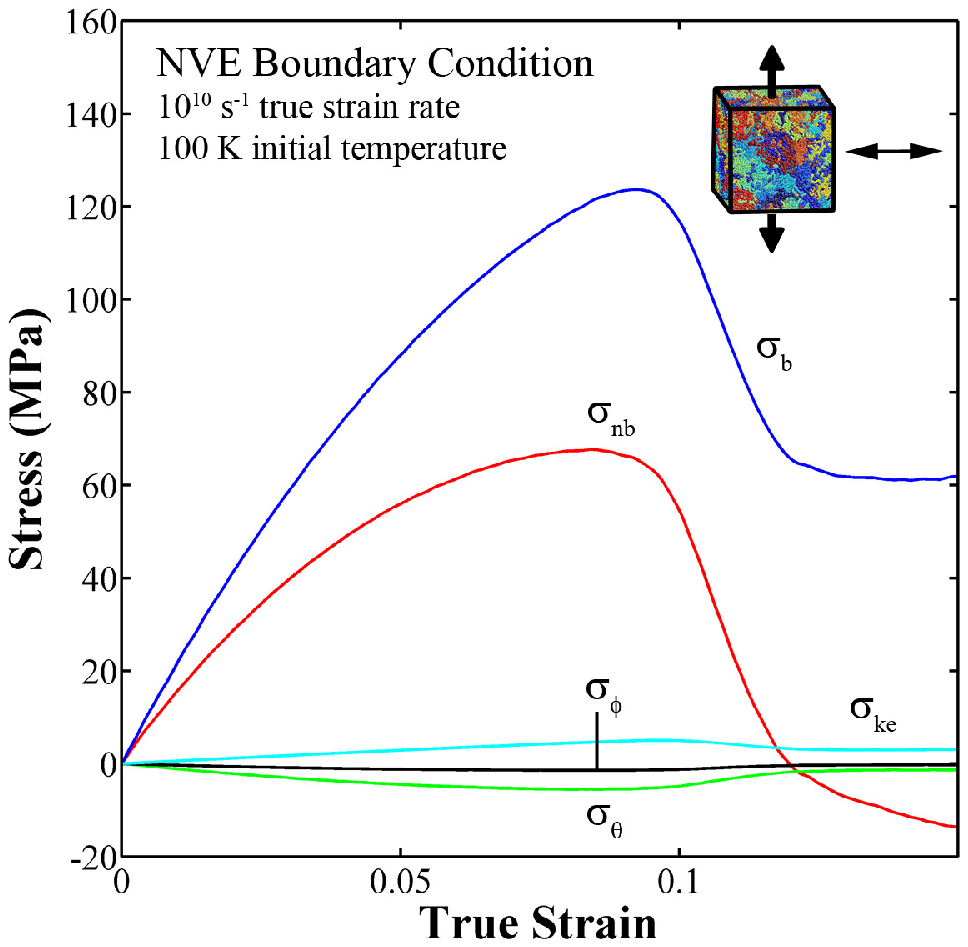} \\
		(c) & (d)
		\end{tabular}
\caption{Normalized stress evolution in the lateral directions as a function of true strain for polyethylene for the (a) NPT, (b) NPH, (c) NVT, and (d) NVE conditions. These curves have been normalized by subtracting the stress values for each component at $\epsilon=0$.}
  \label{fig:fig5}
\end{figure*}

Fig.~\ref{fig:fig5} shows the stress component evolution as a function of strain in the direction perpendicular to the loading direction for all boundary conditions.  Similar to Fig.~\ref{fig:fig3b}, these values have been normalized, too.  A number of trends are very similar to Fig.~\ref{fig:fig3b}.  However, the non-bonded stress component decreases with increasing strain for the NPT and NPH boundary conditions, as shown in Fig.~\ref{fig:fig6}.  Notice that to enforce the zero pressure condition on the boundary, the change in the stress components due to bond stretching and non-bonded Van der Waal's interactions are nearly equal and opposite in magnitude.  The evolution of the stress components for the NVT and NVE boundary conditions are almost identical to that in the direction parallel to the loading direction, as expected due to the similar stress-strain responses (Fig.~\ref{fig:fig3}) and the triaxial state of stress.

\subsection{Internal structure evolution}

In this section, the evolution of the internal structure of the polyethylene system is analyzed.  For the sake of brevity, only the NPT and NPH boundary conditions are examined herein.  For interest in the zero strain boundary conditions (NVT, NVE), we refer the reader to recent articles focusing on the cavitation in amorphous polymers~\cite{Est2011,Mak2011}.

\subsubsection{Bond length and bond angle evolution}

\begin{figure*}[t]
  \centering
		\begin{tabular}{cc}
		\includegraphics[width=3in,angle=0]{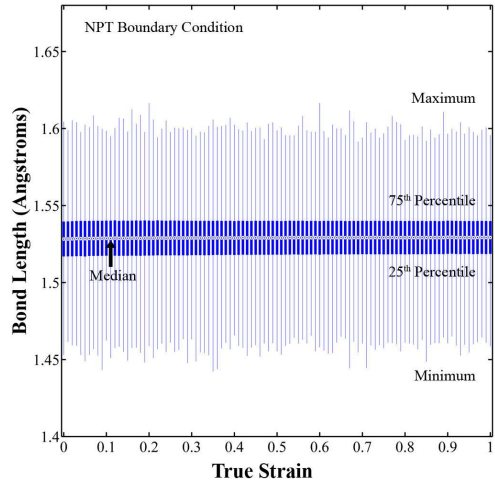} & \includegraphics[width=3in,angle=0]{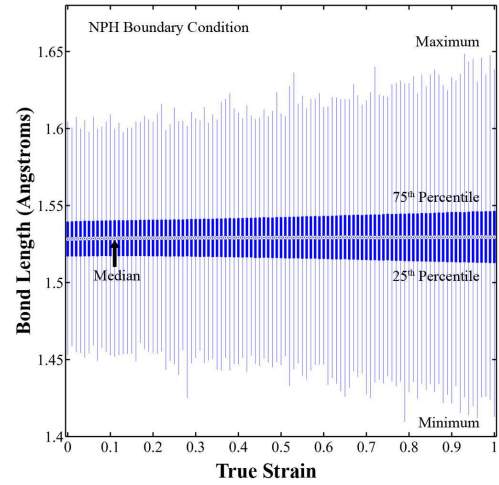}\\
		(a) & (b) \\
		\includegraphics[width=3in,angle=0]{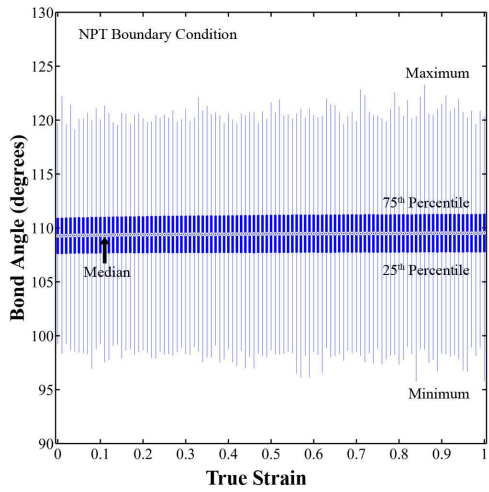} & \includegraphics[width=3in,angle=0]{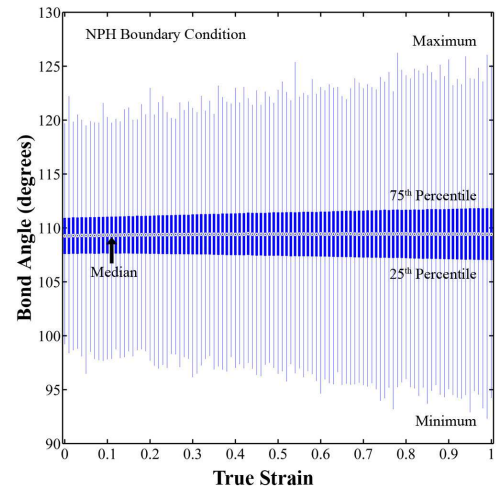} \\
		(c) & (d)
		\end{tabular}
	\caption{Evolution of the (a,b) bond length and (c,d) bond angle distributions as a function of strain for the (a,c) NPT and (b,d) NPH conditions.  While the mean values of the distributions do not significantly change, the distribution width increases as a result of no heat dissipation. }
  \label{fig:fig12}
\end{figure*}

The evolution of the bond length and bond angle distributions as a function of strain may shed light on deformation in amorphous polymer systems.  Fig.~\ref{fig:fig12} shows the evolution of these distributions with a box plot, a concise representation of the $10^7$ data points for each plot.   For each strain level (each box), the central mark is the median, the edges of the filled box are the 25$^{th}$ and 75$^{th}$ percentiles, and the lines extend to the most extreme data points (minimum and maximum), as shown in Fig.~\ref{fig:fig12}(a).  Fig.~\ref{fig:fig12}(a) and (c) show the evolution of the bond length and bond angle distributions, respectively, for the NPT boundary conditions, while Fig.~\ref{fig:fig12}(b) and (d) show the same for the NPH boundary conditions.  The influence of heat dissipation at the boundary has a minimal effect on the bond length and bond angle distributions.  The increase in temperature for the NPH conditions results in a slight broadening of the distribution (as evidenced by the change in the 25$^{th}$ and 75$^{th}$ percentiles as well as the minimum and maximum values), but the median shows almost no change for the symmetric distributions.  Recall that the bond length and bond angle have non-negligible components of the macroscopic stress, though.  To understand their influence on stress, it is also important to quantify the re-orientation of the polymer chain segments.

\subsubsection{Chain orientation evolution}

The chain orientation as a function of strain is also an important microstructure response of the polymer chain segments during deformation.  The chain orientation parameter used here is a second-order Legendre polynomial of ${cos}\theta$, as used for characterizing crystallinity and alignment of polyethylene~\cite{Lav2003,Lee2011}.  First, the local chain orientation at each atom $i$ was computed from the vector connecting nearest neighbor atoms: $e_i=\left(r_{i+1}-r_{i-1}\right)/\left|r_{i+1}-r_{i-1}\right|$.  The orientation $\theta$ was then calculated via $\theta={e_i \cdot e_x}/{\left|e_i\right|\left|e_x\right|}$ and the alignment of the chain segments in the direction of applied stress was calculated using 

\begin{equation}
	P_{2x}=\frac{3}{2}\left\langle cos^2\theta\right\rangle-\frac{1}{2}
\end{equation}

\noindent where $e_x$ is the vector in the direction of applied stress (or orthogonal to the applied stress direction).  In this manner, a value of 1 signifies pure chain segment alignment with the corresponding direction, a value of -0.5 signifies that the chain segment is orthogonal to the corresponding direction, and a value of 0 signifies a randomly oriented sample ($\left\langle cos\theta \right\rangle$)~\cite{Bow2002}.  

\begin{figure*}[t]
  \centering
	\includegraphics[width=4in,angle=0]{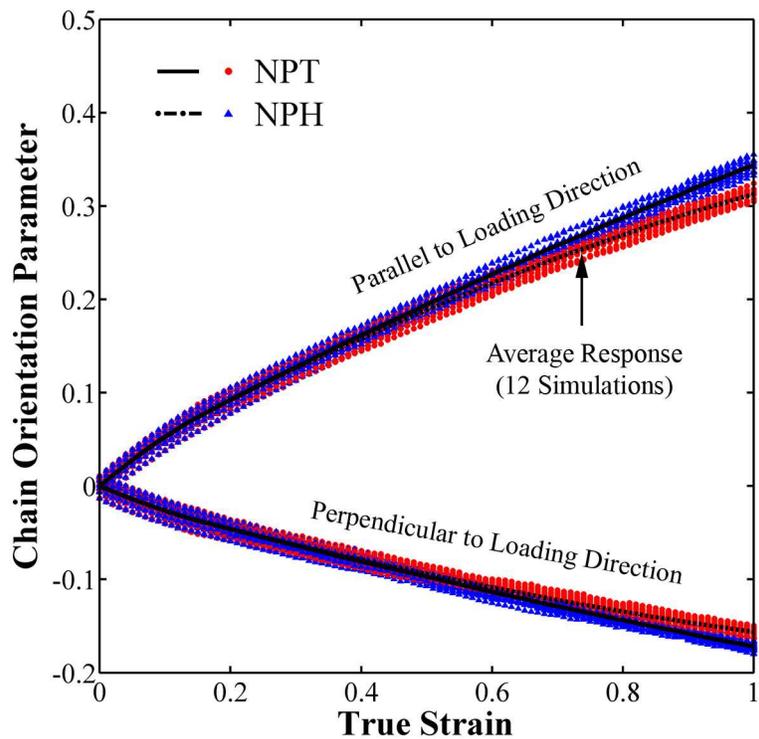}
	\caption{Chain orientation evolution as a function of true strain for polyethylene.  The red circles and blue triangles show the individual responses for twelve simulations ran for each boundary condition. The lack of heat dissipation results in faster re-orientation of the chain segments in the loading direction.}
  \label{fig:fig8}
\end{figure*}

Fig.~\ref{fig:fig8} shows how the average chain orientation parameter, $P_{2x}$, evolves as a function of strain.  On average, the chain segments tend to increase their alignment with the loading direction with increasing strain while decreasing their alignment in the transverse directions.  Very little scatter is observed between the 12 simulations for each boundary condition.  At large strains, polyethylene chains within the constant enthalpy (increasing temperature) NPH boundary condition align more with the tensile direction than the constant temperature NPT condition, as would be expected.  Interestingly, although the bond length and bond angle distributions do not change significantly, the re-orientation and alignment of the polymer chain segments can create significant directional stress components.

\subsubsection{Chain dihedral conformation evolution}

\begin{figure*}[t]
  \centering
		\includegraphics[width=\textwidth,angle=0]{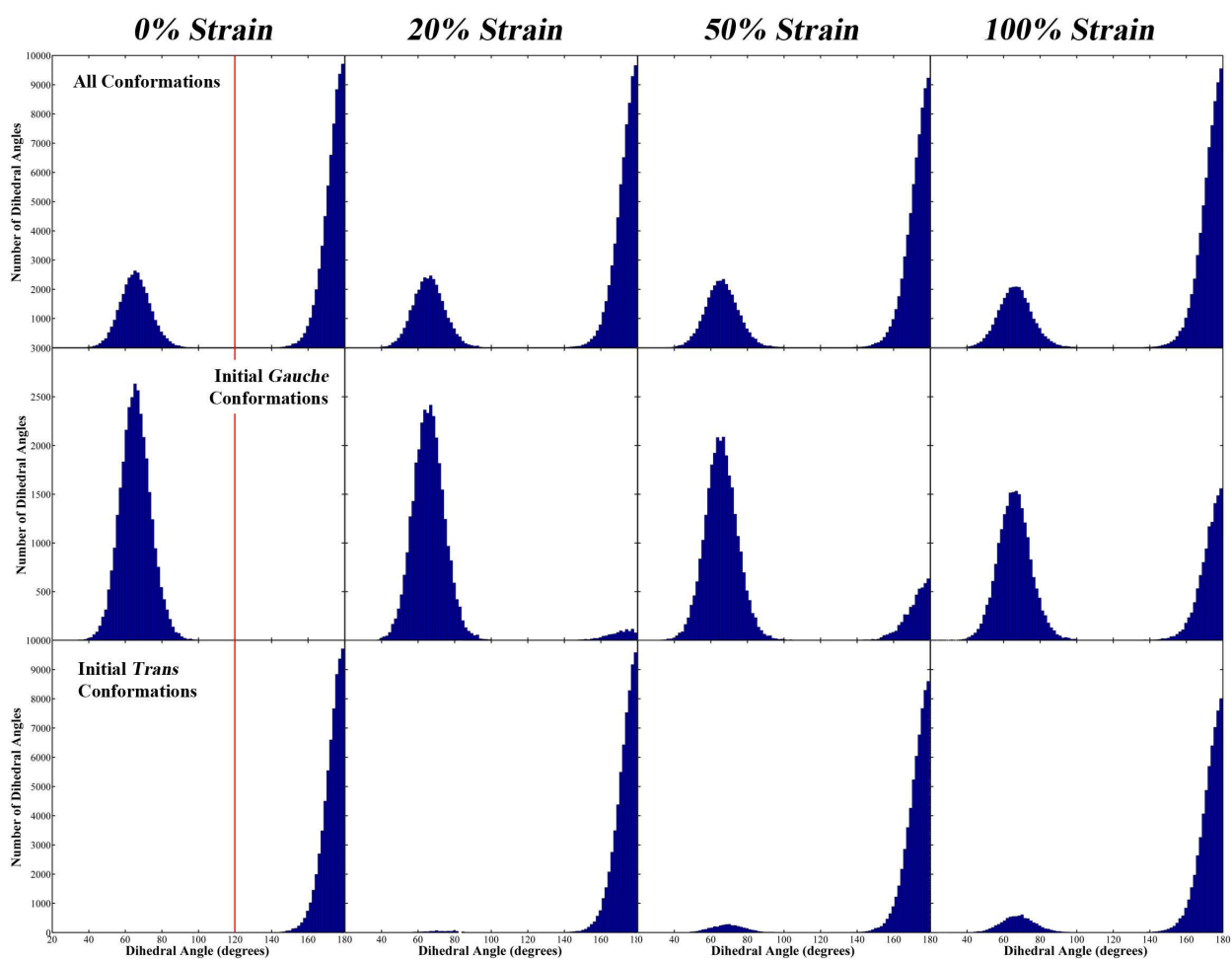} 
		\caption{Evolution of the dihedral angle distribution as a function of strain (0\%, 20\%, 50\%, 100\%) for the NPT condition.  The change in the distribution for all conformations (top) is broken into two distributions based on their initial conformation at 0\% strain: \textit{gauche} and \textit{trans} distributions.  The rate of change between these conformations controls the overall distribution.}
  \label{fig:fig13}
\end{figure*}

The change in the dihedral distribution from metastable \textit{gauche} conformations to stable \textit{trans} conformations as a function of strain is also important for deformation in polymer systems.  Fig.~\ref{fig:fig13} shows the evolution of the dihedral angle distributions for a few strain levels (0\%, 20\%, 50\%, 100\%) with the NPT boundary condition.  The top row tracks the evolution of all segments, the middle row tracks the evolution of those segments that are initially in a \textit{gauche} conformation, and the bottom row tracks the segments that are initially in a \textit{trans} conformation.  The  \textit{gauche} and \textit{trans} conformations were calculated using a dihedral angle threshold of 120$^\circ$.  Some of the \textit{gauche} conformations in the zero strain structure transform to \textit{trans} conformations, and vice versa.  The rate of this transformation is important for accommodating strain and can contribute to chain alignment with the direction of applied stress as well.

\begin{figure*}[!t]
  \centering
		\begin{tabular}{c}
	\includegraphics[width=3in,angle=0]{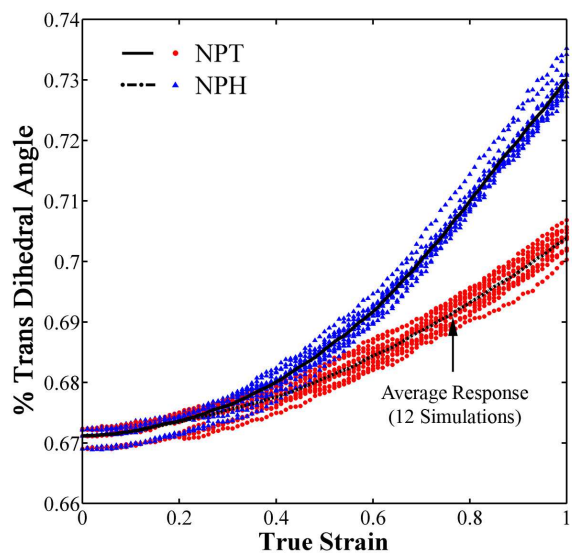} \\
	(a) \\
	\includegraphics[width=3in,angle=0]{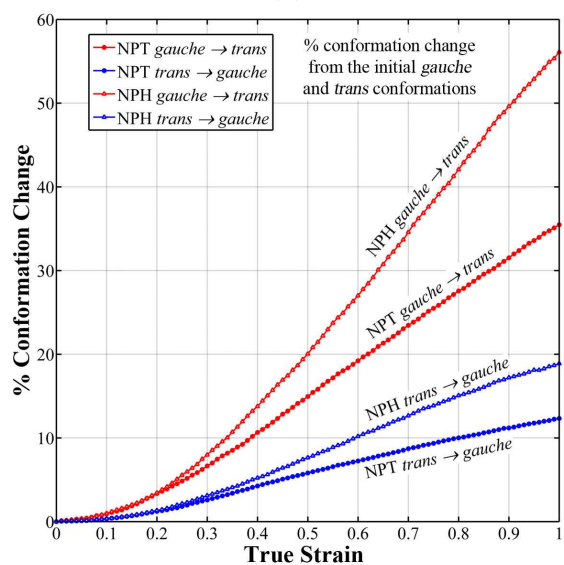} \\ 
  (b)
		\end{tabular}
	\caption{Chain rotation evolution as measured by the change in (a) the percentage of \textit{trans} dihedral angles and (b) the conformation changes between \textit{gauche} and \textit{trans} distributions as a function of true strain for polyethylene.  The red circles and blue triangles show the individual responses for twelve simulations ran for each boundary condition.  The lack of heat dissipation results in a higher rate of change between conformations that results in an increased percentage of \textit{trans} conformations, mainly in the strain hardening regime.}
  \label{fig:fig7}
\end{figure*}

Fig.~\ref{fig:fig7} shows how both (a) the rate of transformation (\% conformation change) and (b) the percent \textit{trans} conformations evolve as a function of strain for the NPT and NPH boundary conditions.  Fig.~\ref{fig:fig7}(a) shows that the rate of change of the conformations from \textit{gauche}$\rightarrow$\textit{trans} and \textit{trans}$\rightarrow$\textit{gauche} is increased significantly for the NPH condition.  Additionally, the rate of change of conformations is approximately 2.5-3.0 times greater for the \textit{gauche}$\rightarrow$\textit{trans} transformation than the \textit{trans}$\rightarrow$\textit{gauche} tranformation for $\epsilon\ge{20}\%$.  Fig.~\ref{fig:fig7}(b) shows the net effect on the \% \textit{trans} conformations.  Along with the average behavior, the results of the twelve individual simulations for each boundary condition are also plotted to show the range of data.  The 10$^{10}$ s$^{-1}$ shows very little change in the percentage of \textit{trans} conformations in the elastic regime, but this percentage increases with increasing strain in the strain softening and strain hardening regimes.  The percentage of \textit{trans} dihedral angles increases at a much faster rate in the constant enthalpy (increasing temperature) NPH condition as compared to the constant temperature NPT condition.  As shown here, the increased temperature allows the transformation between \textit{gauche} and \textit{trans} to be more easily overcome, which enables a higher \textit{gauche}$\rightarrow$\textit{trans} rate of change to align chain segments with the direction of loading.

\subsubsection{Chain entanglement evolution}

The chain entanglement evolution is also important for understanding deformation in amorphous polymer systems.  In general, fracture of many polymer materials is expected to occur due to chain disentanglement rather than chain scission, because of the strong carbon-carbon bonds in the polymer backbone.  Here, the geometric technique of Yashiro et al.~\cite{Yas2003}~was used to calculate the chain entanglement.  This technique first creates two vectors which emanate from each atom to neighboring atoms that are separated by 10 atoms on the same chain, i.e., one vector that connects atom $i$ with atom ($i-10$) and one vector that connects atom $i$ to atom ($i+10$).  The angle between these two vectors is calculated for each applicable atom and a threshold parameter of 90$^\circ$ is used to specify if the atom is classified as entangled or not.  The atoms classified as entangled via this technique are considered to be constrained by neighboring chains.  The number of atoms classified as entangled is then divided by the total number of applicable atoms to give a normalized ``entanglement parameter'' that can be used to compare the different systems.  This entanglement parameter actually represents the percent of entangled atoms within the system and can be appropriately scaled by a constant to obtain the entanglement density.  We refrain from using an entanglement density here, since this measure is dependent on the separation distance used for creating the two vectors and also the threshold parameter.  However, these parameters simply affect the scaling of the y-axis and not the general trend.  

\begin{figure*}[!t]
  \centering
	\includegraphics[width=4in,angle=0]{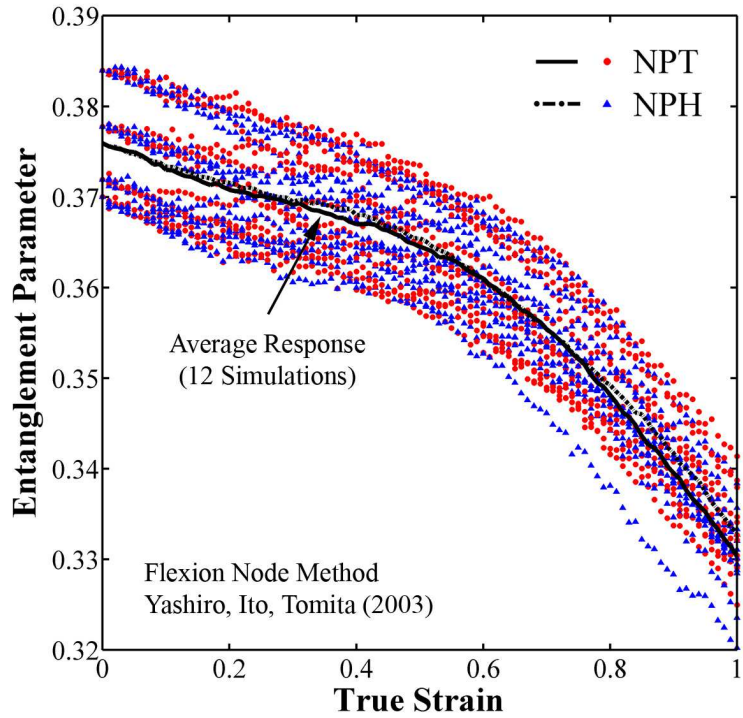}
	\caption{Chain entanglement evolution as a function of true strain for polyethylene.  The chain entanglement is measured by the flexion node method of Yashiro et al.~\cite{Yas2003}.  The chain entanglement parameter decreases at a faster rate in the strain hardening region with a low sensitivity to the boundary condition.}
  \label{fig:fig9}
\end{figure*}

Fig.~\ref{fig:fig9} shows the evolution of the entanglement parameter as a function of strain.  There is very little difference between the entanglement parameter for the NPT and NPH boundary conditions. The entanglement parameter decreases monotonically with increasing strain.  This result agrees with previous work of Tomita~\cite{Tom2000} and Shepherd et al.~\cite{She2006}, which assume that the entanglement density decreases with an increase of the strain level at constant strain rate and temperature.  However, the rate of entanglement decrease with increasing strain is not constant.  At $\approx$50\% strain, the entanglement parameter transitions from a region of slow linear decrease to a region of faster linear decrease.  Based on previous findings herein, it is speculated that the faster rate of disentanglement is due to the reorientation of chains as well as conformation changes, both of which allow entangled chain segments to disentangle.     

\section{Conclusions}

The heat and pressure dissipation was studied to determine its impact on the deformation of amorphous polyethylene using molecular dynamics simulations.  While the tensile direction was deformed at a constant strain rate, the lateral boundaries were controlled via a zero strain condition (NVT, NVE) or a zero stress condition (NPT, NPH) and either a thermostat was used to maintain constant temperature during deformation (NPT, NVT) or a constant enthalpy/energy condition was employed where the temperature increased during deformation (NPH, NVE).  The energy and stress partitioning was examined as well as the evolution of the internal structure of the amorphous polymer.  The following conclusions can be drawn from this study:

\begin{enumerate}
	\item The large increase in temperature as a function of strain observed for the non-thermostat conditions (NPH, NVE) resulted in minor changes to the stress-strain behavior, primarily at higher strains in the strain hardening region (Fig.~\ref{fig:fig3}).  However, the lack of pressure dissipation caused a high triaxial state of stress in the deformed PE system, which results in void nucleation, void growth, and heterogeneous deformation (Fig.~\ref{fig:fig10}).  
	\item Analysis of the energy partitioning shows that a majority of the internal energy increase during deformation is due to the non-bonded Van der Waal's interactions for all of the boundary conditions used in this study (Fig.~\ref{fig:fig3a}).  To a lesser extent, the energies attributed to bond stretching (bond length), bond bending (bond angles), and bond torsion (dihedral angles) also contributed during deformation.  The kinetic energy component is a significant portion of the total energy at higher strains ($>$0.2\%) for the constant enthalpy NPH boundary condition (increasing energy).     
	\item Analysis of the stress partitioning shows that the macroscopic stress is a mainly balance between the `tensile' non-bonded Van der Waal's interactions and the `compressive' bond stretching term (Fig.~\ref{fig:fig6}).  During tensile deformation, the stresses associated with the Van der Waal's interactions and bond stretching increase in the tensile direction for all cases considered in this study.  Interestingly, the characteristic yield stress peak prior to strain softening and strain hardening is clearly due to the Van der Waal's interactions between chains (Fig.~\ref{fig:fig3b}).  To maintain the zero stress condition for the pressure dissipation cases (NPT, NPH), both the `tensile' non-bonded stress and the `compressive' bond stretching stress relax toward zero stress (i.e., bonded stress increases and non-bonded stress decreases).
	\item The following analysis concentrates on the pressure dissipation boundary conditions (NPT, NPH).  The bond length and bond angle distributions do not change significantly as a function of strain.  For the case of increased temperature, there is a slight broadening of the symmetric distributions but no increase or descrease of the distribution.  Rather, the large contribution of bond stretching and bond bending to the macroscopic stress are due to the re-orientation of the polymer chain segments in the direction of loading (Fig.~\ref{fig:fig8}), which results in a larger force being resolved in the direction of loading.  The increase in temperature for the NPH case results in a faster rate of re-orientation of the polymer chain segments, which results in a higher stress in the direction of loading, but this is offset by the stress contribution of the kinetic energy component in molecular dynamics simulations.
	\item The conformational changes from \textit{gauche} to \textit{trans}, and vice versa, result in non-negligible changes to the energy, but very little change to the macroscopic stress.  In other words, conformational changes contribute more to the deformation kinematics (i.e., texture) and the ability of the polymer chains to align in the direction of loading than to the actual load carrying capacity.  Fig.~\ref{fig:fig13} and Fig.~\ref{fig:fig7} show that the change in the dihedral angle distribution as a function of strain is a result of both \textit{gauche}$\rightarrow$\textit{trans} and \textit{trans}$\rightarrow$\textit{gauche} changes at different rates ($\approx$2.5-3.0 times faster for \textit{gauche}$\rightarrow$\textit{trans}).  The increase in temperature for the NPH case results in a faster rate of change for the  \textit{gauche} and \textit{trans} conformations.
	\item The chain entanglement decreases as a function of strain with a large rate of decrease in the strain hardening regime ($\epsilon>$0.5). 	The entanglement is not affected very much by the thermostat (NPT versus NPH) but does show some variation between the twelve instantiations studied for each condition (Fig.~\ref{fig:fig9}).
\end{enumerate}

\section*{Acknowledgments}

This work was performed at the Center for Advanced Vehicular Systems (CAVS) at Mississippi State University.  This material is based upon work supported by the U.S. Army TACOM Life Cycle Command under Contract No.~W56HZV-08-C-0236, through a subcontract with Mississippi State University, and was performed for the Simulation Based Reliability and Safety (SimBRS) research program. Reference herein to any specific commercial company, product, process, or service by trade name, trademark, manufacturer, or otherwise, does not necessarily constitute or imply its endorsement, recommendation, or favoring by the United States Government or the Department of the Army (DoA). The opinions of the authors expressed herein do not necessarily state or reflect those of the United States Government or the DoA, and shall not be used for advertising or product endorsement purposes.

\bibliographystyle{ieeetr}

\clearpage
\newpage

\begin{table}[t]
	\centering
		\begin{tabular}{cc}
		\hline
			Parameters & Values \\
		\hline
			$K_b$ & 350 kcal/mol \\
			$r_0$ & 1.53 \AA\ \\
			$K_\theta$ & 60 kcal/mol/rad$^2$ \\
			$\theta_0$ & 1.911 rad \\
			$C_0$ & 1.736 kcal/mol \\
			$C_1$ & -4.490 kcal/mol \\
			$C_2$ & 0.776 kcal/mol \\
			$C_3$ & 6.990 kcal/mol \\
			$\sigma_0$ & 4.01 \AA\ \\			
			$\epsilon_0$ & 0.112 kcal/mol \\
		\hline
		\end{tabular}
	\caption{Interatomic potential parameters}
	\label{tab:InteratomicPotentialParameters}
\end{table}


\end{document}